\documentclass[entropy,article,submit,moreauthors,pdftex]{Definitions/mdpi} 
\usepackage[]{amsmath}
\usepackage{comment}
\usepackage{physics}
\usepackage{bbold}
\usepackage{amsmath}
\usepackage{amsfonts}
\usepackage{cases}
\usepackage{xspace}
\usepackage{psfrag}
\usepackage{subfig}
\usepackage{color}
\usepackage{xcolor}

\newcommand{\ave}[1]{\left\langle #1 \right\rangle}

\newcommand{\plaind}{\mathrm{d}}
\newcommand{\dint}[1]{\mathchoice{\!\plaind#1\,}{\!\plaind#1\,}{\!\plaind#1\,}{\!\plaind#1\,}}

\usepackage{dsfont}

\newcommand{\canetset}[1]{{\mathchoice {\hbox{$\sf\textstyle #1\kern-0.4em #1$}}
{\hbox{$\sf\textstyle #1\kern-0.4em #1$}}
{\hbox{$\sf\scriptstyle #1\kern-0.3em #1$}}
{\hbox{$\sf\scriptscriptstyle #1\kern-0.2em #1$}}}}

\def\nbZ{{\mathchoice {\hbox{$\sf\textstyle Z\kern-0.4em Z$}}
{\hbox{$\sf\textstyle Z\kern-0.4em Z$}}
{\hbox{$\sf\scriptstyle Z\kern-0.3em Z$}}
{\hbox{$\sf\scriptscriptstyle Z\kern-0.2em Z$}}}}

\newcommand{\gpvec}[1]{\mathbf{#1}}

\newcommand{\mvec}{\gpvec{m}}
\newcommand{\nvec}{\gpvec{n}}

\newcommand{\xvec}{\gpvec{x}}

\newcommand{\zvec}{\gpvec{z}}

\newcommand{\Fvec}{\gpvec{F}}

\newcommand{\Pvec}{\gpvec{P}}

\newcommand{\IC}{\mathcal{I}}

\newcommand{\OC}{\mathcal{O}}

\newcommand{\ZC}{\mathcal{Z}}

\newcommand{\half}{\mathchoice{\frac{1}{2}}{(1/2)}{\frac{1}{2}}{(1/2)}}

\renewcommand{\exp}[1]{\mathchoice{\mathrm{e}^{#1}}{\operatorname{exp}\left(#1\right)}{\operatorname{exp}\left(#1\right)}{\operatorname{exp}\left(#1\right)}}

\newcommand{\elabel}[1]{\label{eq:#1}}
\newcommand{\eref}[1]{(\ref{eq:#1})}
\newcommand{\Eref}[1]{Eq.~(\ref{eq:#1})}
\newcommand{\Erefs}[1]{Eqs.~(\ref{eq:#1})}

\newcommand{\sref}[1]{Sec.~\ref{sec:#1}}
\newcommand{\Sref}[1]{Section~\ref{sec:#1}}

\newcommand{\flabel}[1]{\label{fig:#1}}
\newcommand{\fref}[1]{Fig.~\ref{fig:#1}}

\newcommand{\latin}[1]{{\it #1}}
\newcommand{\ie}{\latin{i.e.}\@\xspace}

\newcommand{\rcomment}[1]{{\textcolor{cyan}{\bf RGM: #1}}}

\newlength \standardfigwidth
\DeclareMathAlphabet{\matheub}{U}{eur}{m}{n}

\newcounter{exercise}
{\addtocounter{exercise}{1}\begin{center}\begin{minipage}{0.8\linewidth}\textbf{Exercise
\arabic{exercise}:}\begin{itshape}}
{\end{itshape}\end{minipage}\end{center}}

\makeatletter
\newcommand{\creat}[3][]{\@ifempty{#1}{#2^{\dagger}}{\left(#2^{\dagger}\right)^{#1}}\@ifempty{#3}{}{\!(#3)}}

\newcommand{\creatDoi}[3][]{\@ifempty{#1}{\tilde{#2}}{\left(\tilde{#2}\right)^{#1}}\@ifempty{#3}{}{(#3)}}

\newcommand{\annih}[3][]{#2\@ifempty{#1}{}{^{#1}}\@ifempty{#3}{}{(#3)}}

\makeatother

% ...

\newlength{\bibmarkkeyAleft}

\newlength{\bibmarkkeyBleft}

\newlength{\bibmarkkeyCleft}

\newlength{\bibmarkkeyDleft}

\usepackage{natbib}

\newcommand{\entropyProduction}{\dot{S}_i}

\newcommand{\entropyFlow}{\dot{S}_e}

\newcommand{\Deltat}{\Delta t}

\newcommand{\latticespacing}{a}
\newcommand{\drift}{v}
\newcommand{\weight}{z}

\preto{\abstractkeywords}{\nolinenumbers}

\firstpage{1} 
\pubvolume{xx}
\issuenum{1}
\articlenumber{5}
\pubyear{2019}
\copyrightyear{2019}
\history{Received: date; Accepted: date; Published: date}

\Title{Entropy production in exactly solvable systems}

\Author{
Luca Cocconi$^{1,2,3,\dagger}$\orcidA{},
Rosalba Garcia-Millan$^{1,2,4,\dagger}$\orcidB{},
Zigan Zhen$^{1,2}$\orcidC{},
Bianca Buturca$^{5}$\orcidD{},
Gunnar Pruessner$^{1,2}$\orcidE{}*}

\AuthorNames{Luca Cocconi, Rosalba Garcia-Millan, Zigan Zhen, Bianca Buturca and Gunnar Pruessner}

\address{%
$^{1}$ \quad Department of Mathematics, Imperial College London, 180 Queen's Gate, London SW7 2AZ, UK\\
$^{2}$ \quad Centre for Complexity Science, Imperial College London\\
$^{3}$ \quad Francis Crick Institute, 1 Midland Rd, London NW1 1AT, UK\\
$^{4}$ \quad DAMTP, Centre for Mathematical Sciences, University of Cambridge, Wilberforce Road, Cambridge CB3 0WA, UK\\
$^{5}$ \quad Department of Physics, Imperial College London, Exhibition Road, London SW7 2AZ, UK\\
$\dagger$ \quad These authors have contributed equally to this work.
}

\corres{Correspondence: g.pruessner@imperial.ac.uk}

\abstract{
The rate of entropy production by a stochastic process quantifies how far it is from thermodynamic equilibrium. Equivalently, entropy production captures the degree to which detailed balance and time-reversal symmetry are broken.
Despite abundant references to entropy production in the literature and its many applications in the study of non-equilibrium stochastic particle systems, a comprehensive list of typical examples illustrating the fundamentals of entropy production is lacking.
Here, we present a brief, self-contained review of entropy production and calculate it from first principles in a catalogue of exactly solvable setups, encompassing both discrete- and continuous-state Markov processes, as well as single- and multiple-particle systems. 
The examples covered in this work provide a stepping stone for further studies on entropy production of more complex systems, such as many-particle active matter, as well as a benchmark for the development of alternative mathematical formalisms.
}


\begin{document}


\section{Introduction}
 Stochastic thermodynamics has progressively evolved into an essential tool in the study of non-equilibrium systems as it connects the quantities of interest in traditional thermodynamics, such as work, heat and entropy, to the properties of microscopically-resolved fluctuating trajectories \cite{seifert_stochastic_2012,jiang_mathematical_2004,seifert_stochastic_2018}. The possibility of equipping stochastic processes with a consistent thermodynamic and information-theoretic interpretation has resulted in a number of fascinating works, with the interface between mathematical physics and the biological sciences proving to be a particularly fertile ground for new insights (e.g.\ \cite{barato_efficiency_2014,lan_information_2016,cao_free-energy_2015,schmiedl_stochastic_2007,pietzonka_autonomous_2019}). The fact that most of the applications live on the small scale is not surprising, since it is precisely at the microscopic scale that  fluctuations start to play a non-negligible r{\^o}le.

 The concept of entropy and, more specifically, entropy production has attracted particular interest, as a consequence of the quantitative handle it provides on the distinction between equilibrium systems, passive systems relaxing to equilibrium and genuinely non-equilibrium, `active' systems. While there exist multiple routes to the mathematical formulation of entropy production \cite{schnakenberg_network_1976,maes_fluctuation_1999,gaspard_time-reversed_2004,Seifert:2005,nardini_entropy_2017,landi_entropy_2013}, the underlying physical picture is consistent: the entropy production associated with an ensemble of stochastic trajectories quantifies the degree of certainty with which we can assert that a particular event originates from a given stochastic process or from its suitably defined conjugate (usually, its time-reverse). When averaged over long times (or over an ensemble), a non-vanishing entropy production signals time-reversal symmetry breaking at the microscopic scale. This implies, at least for Markovian systems, the existence of steady-state probability currents in the state space, which change sign under time-reversal. When a thermodynamically consistent description is available, the average rate of entropy production can be related to the rate of energy or information exchange between the system, the heat bath(s) it is connected to, and any other thermodynamic entity involved in the dynamics, such as a measuring device \cite{munakata_entropy_2014,loos_heat_2019,ouldridge_power_2018}. Whilst the rate of energy dissipation is of immediate interest since it captures how `costly' it is to sustain specific dynamics (e.g. the metabolism sustaining the development of an organism \cite{noauthor_heat_nodate,song_energy_2019}), entropy production has also been found to  relate non-trivially to the efficiency and precision of the corresponding process via uncertainty relations \cite{horowitz_thermodynamic_2020,seifert_stochastic_2018}. Entropy production along fluctuating trajectories also plays a fundamental r{\^o}le in the formulation of various fluctuation theorems \cite{Seifert:2005}. 

 Given the recent interest in stochastic thermodynamics and entropy production in particular, as well as the increasing number of mathematical techniques implemented for the quantification of the latter, it is essential to have available a few, well-understood reference systems, for which exact results are known. These can play the r{\^o}le of benchmarks for new techniques, while helping neophytes to develop intuition. In this work, we will present results exclusively in the framework proposed by Gaspard \cite{gaspard_time-reversed_2004}, specifically in the form of \Erefs{external_en_prod_and_internal_en_prod},
(\ref{eq:continuum_en_prod}) and (\ref{eq:continuum_en_flow}), which we review and contextualise by deriving them via different routes
in Section \ref{s:review_en}. 
In Section \ref{S:systems} we begin the analysis with processes in discrete state space (Sections \ref{sec:sec2state}-\ref{sec:RWring}), and subsequently extend it to the continuous case (Sections \ref{sec:DDparticle}-\ref{sec:Bpartring}). Finally, in sections \ref{sec:RnTring} and \ref{sec:switchdiff} we consider processes that involve both discrete and continuous degrees of freedom. Time is taken as a continuous variable throughout. 

\section{Brief review of entropy production \label{s:review_en}}

\noindent \textit{Entropy production of jump processes.} The concept of time-dependent informational entropy associated with a given ensemble of stochastic processes was first introduced by Shannon \cite{shannon_mathematical_nodate}. For an arbitrary probability mass function $P_n(t)$ of time $t$ 
over a discrete set of states $n \in \Omega$, the Shannon entropy is defined as 
\begin{equation}
    \label{eq:entropy}
    S(t) = - \sum_n P_n(t) \ln P_n(t)
\end{equation}
with the convention henceforth of $x\ln x=0$ for $x=0$. It quantifies the inherent degree of uncertainty about the state of a process. In the microcanonical ensemble $P_n$ is constant in $t$ and $n$ and upon providing an entropy scale in the form of the Boltzmann constant $k_B$, Shannon's entropy reduces to that of traditional thermodynamics given by Boltzmann's $S=k_B \ln |\Omega|$, where $|\Omega|=1/P_n$ is the cardinality of $\Omega$. In Markovian systems, the probability $P_n(t)$ depends on $n$ and evolves in time $t$ according to the master equation
\begin{equation}
\label{eq:markov_master_eq}
    \dot{P}_n(t) = \sum_{m} P_m(t) w_{m  n} - P_n(t) w_{n m} 
\end{equation}
with non-negative transition rates $w_{m n}$ from state $m$ to state $n \neq m$.
\Eref{markov_master_eq} reduces to $\dot{P}_n(t) = \sum_m P_m(t) w_{mn}$ by imposing the Markov condition $\sum_m w_{nm} = 0$, which we will use in the following. For simplicity we will restrict ourselves to time-independent rates $w_{nm}$ but as far as the following discussion is concerned, generalising to time-dependent rates is a matter of replacing $w_{nm}$ by $w_{nm}(t)$. The rate of change of entropy for a continuous time jump process can be derived by differentiating $S(t)$ in \Eref{entropy} with respect to time and substituting (\ref{eq:markov_master_eq}) into the resulting expression \cite{gaspard_time-reversed_2004,Esposito:2010}, thus obtaining
\begin{equation}
\label{eq:time_der_entr}
\dot{S}(t) 
= - \sum_{m,n} P_m(t) w_{mn} \ln \left( P_n(t) \right) 
= \sum_{m,n} P_n(t) w_{nm} \ln \left( \frac{P_n(t)}{P_m(t)} \right) 
= \entropyFlow(t) + \entropyProduction(t)
\end{equation}
where we define
\begin{subequations}
\elabel{external_en_prod_and_internal_en_prod}
\begin{align}
\entropyFlow(t)
&= - \half \sum_{m,n} \left( P_n(t) w_{nm} - P_m(t) w_{mn} \right) \ln \left( \frac{w_{nm}}{w_{mn}} \right) 
\label{eq:external_en_prod}\\
&= - \sum_{m,n} P_n(t) w_{nm} \ln \left( \frac{w_{nm}}{w_{mn}} \right) 
= - \sum_{m,n}  \left( P_n(t) w_{nm} - P_m(t) w_{mn} \right) \ln \left( \frac{w_{nm}}{w_{0}} \right) \nonumber \\ 
\entropyProduction(t)
&= \half \sum_{m,n} \left( P_n(t) w_{nm} - P_m(t) w_{mn} \right) \ln \left( \frac{P_n(t) w_{nm}}{P_m(t) w_{mn}} \right)
\label{eq:internal_en_prod} \\
&= \sum_{m,n} P_n(t) w_{nm} \ln \left( \frac{P_n(t) w_{nm}}{P_m(t) w_{mn}} \right) 
= \sum_{m,n} (P_n(t) w_{nm} - P_m(t) w_{mn}) \ln \left( \frac{P_n(t) w_{nm}}{w_{0}} \right)\nonumber  
\end{align}
\end{subequations}
with arbitrary positive rate $w_0$ 
to restore dimensional consistency, that cancel trivially.
Here we follow the convention \cite{seifert_stochastic_2012} to split the rate of entropy change into two contributions: the first, Eq.~(\ref{eq:external_en_prod}), commonly referred to as ``external" entropy production or entropy flow, is denoted by $\entropyFlow$. It contains a factor $\ln(w_{nm}/w_{mn})$ corresponding, for systems satisfying local detailed balance, to the net change in entropy of the reservoir(s) associated with the system's transition from state $n$ to state $m$. For such thermal systems, $\entropyFlow$ can thus be identified as the rate of entropy production in the environment \cite{lebowitz_spohn_1999,schnakenberg_network_1976}. The second contribution, Eq.~(\ref{eq:internal_en_prod}), termed ``internal" entropy production and denoted by $\entropyProduction$ is non-negative because $(x-y)\ln (x/y) \geq 0$ for any two real, positive $x$, $y$ and using the convention $z \ln z=0$ for $z=0$. The internal entropy production vanishes when the detailed balance condition $P_n(t) w_{nm} = P_m(t) w_{mn}$ is satisfied for all pairs of states. In this sense, a non-vanishing $\entropyProduction$ is the fingerprint of non-equilibrium phenomena. At steady-state, namely when $\dot{P}_n(t)=0$ for all $n$, $\dot{S}(t)$ in Eq.~(\ref{eq:time_der_entr}) vanishes by construction, so that the internal and external contributions to the entropy production cancel each other exactly,
$\dot{S}(t)=\entropyFlow(t)+\entropyProduction(t)=0$,
while they vanish individually only for systems at equilibrium. Equations (\ref{eq:external_en_prod_and_internal_en_prod}) will be used throughout the present work to compute the entropy productions of discrete-state processes.\\

\noindent \textit{Entropy production as a measure of time-reversal-symmetry breaking.} As it turns out, a deeper connection between internal entropy production and time-reversal symmetry breaking can be established \cite{gaspard_time-reversed_2004}. The result, which we re-derive below, identifies $\entropyProduction$ as the relative dynamical entropy (\ie the Kullback-Leibler divergence \cite{kullback_information_1951}) per unit time of the ensemble of forward paths and their time-reversed counterparts. To see this, we first need to define a path $\nvec = (n_0, n_1, \ldots, n_{M})$ as a set of trajectories starting at time $t_0$ and visiting states $n_j$ at successive discrete times $t_j = t_0 + j \tau$ with $j=0,1,\ldots,M$, equally spaced by a time interval $\tau$. For a time-homogeneous Markovian jump process in continuous time, the joint probability of observing a particular path is
\begin{equation}
\label{eq:path_prob}
\mathcal{P}(\nvec;t_0,M\tau) = P_{n_0}(t_0) W(n_0 \to n_1;\tau) W(n_1 \to n_2;\tau) \ldots W(n_{M-1} \to n_M;\tau)
\end{equation}
where $P_{n_0}(t_0)$ is the probability of observing the system in state $n_0$ at time $t_0$, while $W(n_j \to n_{j+1};\tau)$ is the probability that the system is in state $n_{j+1}$ time $\tau$ after being in state $n_j$. This probability can be expressed in terms of the transition rate matrix $w$ with elements $w_{mn}$. It is  $W(n \to m;\tau) = [{\rm exp}(w\tau)]_{nm}$, the matrix elements of the exponential of the matrix $w\tau$ with the Markov condition imposed.
It can be expanded in small $\tau$ as 
\begin{equation}
\label{eq:W_expanded}
W(n \to m;\tau)= \delta_{n,m} + w_{nm} \tau + O(\tau^2) ~,
\end{equation} 
where $\delta_{n,m}$ is the Kronecker-$\delta$ function.
We can now define a dynamical entropy per unit time \cite{shannon_mathematical_nodate} as
\begin{equation}
\label{eq:h_discrete}
 h(t_0,\Delta t) = \lim_{M \to \infty} - \frac{1}{\Delta t} \sum_{n_0,\ldots,n_M} \mathcal{P}(\nvec;t_0,\Delta t) \ln \mathcal{P}(\nvec;t_0,\Delta t) ~.
\end{equation}
where the limit is to be considered a continuous time limit taken at fixed $\Delta t=t_M - t_0 = M\tau$ \cite{Diana_2014}, thus determining the sampling interval $\tau$, and the sum runs over all possible paths $\nvec$. Other than $\tau$, the paths are the only quantity on the right-hand side of Eq.~(\ref{eq:h_discrete}) that depend on $M$. The dynamical entropy $h(t_0,\Delta t)$ may be considered the expectation of $\ln (\mathcal{P}(\nvec;t_0,\Delta t))$ across all paths. Similarly to the static Shannon entropy, the dynamical entropy $h(t_0,\Delta t)$ quantifies the inherent degree of uncertainty about the evolution over a time $\Delta t$ of a process starting at a given time $t_0$. To compare with the dynamics as observed under time-reversal, one introduces the time-reversed path $\nvec^R=(n_M,n_{M-1},\ldots,n_0)$ and thus the time-reversed dynamical entropy per unit time as
\begin{equation}
\label{eq:hR_discrete}
 h^R(t_0,\Delta t) = \lim_{M \to \infty} - \frac{1}{\Delta t} \sum_{n_0,\ldots,n_M} \mathcal{P}(\nvec;t_0,\Delta t) \ln \mathcal{P}(\nvec^R;t_0,\Delta t) ~.
\end{equation}
While similar in spirit to $h(t_0,\Delta t)$, the physical interpretation of $h^R(t_0,\Delta t)$ as the expectation of $\ln (\mathcal{P}(\nvec^R;\Delta t))$ under the forward probability $\mathcal{P}(\nvec;t_0,\Delta t)$ is more convoluted since it involves the forward and the backward paths simultaneously, which have potentially different statistics. However, time-reversal symmetry implies precisely identical statistics of the two ensembles, whence $h(t_0,\Delta t) = h^R(t_0,\Delta t)$. The motivation for introducing $h^R(t_0,\Delta t)$ is that the difference of the two dynamical entropies defined above is a non-negative Kullback-Leibler divergence given by
\begin{equation}
\label{eq:kl_paths}
    h^R(t_0,\Delta t) - h(t_0,\Delta t) = \lim_{M \to \infty} \frac{1}{\Delta t} \sum_\nvec \mathcal{P}(\nvec;t_0,\Delta t) \ln \left( \frac{\mathcal{P}(\nvec;t_0,\Delta t)}{\mathcal{P}(\nvec^R;t_0,\Delta t)} \right) ~.
\end{equation}
Using Eq.~(\ref{eq:path_prob}) in (\ref{eq:kl_paths}) 
with Eq.~(\ref{eq:W_expanded}) provides the expansion
\begin{equation}
\label{eq:kl_final}
    h^R(t_0,\Delta t) - h(t_0,\Delta t) = \sum_{nm} P_n(t_0) w_{nm} \ln \left( \frac{P_n(t_0) w_{nm}}{P_m(t_0) w_{mn}} \right) + \mathcal{O}(\Delta t) ~,
\end{equation}
which is an instantaneous measure of the Kullback-Leibler divergence. 
The limit of 
$h^R(t_0,\Delta t) - h(t_0,\Delta t)$ 
in small $\Delta t$ is finite
and identical to the internal entropy production (\ref{eq:internal_en_prod}) derived above. This result establishes the profound connection between broken detailed balance, Eq.~(\ref{eq:external_en_prod_and_internal_en_prod}), and Kullback-Leibler divergence, Eq.~(\ref{eq:kl_final}), both of which can thus be recognised as fingerprints of non-equilibrium systems. In light of this connection, it might not come as a surprise that the steady-state rate of entropy production is inversely proportional to the minimal time needed to decide on the direction of the arrow of time \cite{Roldan_decision_making2015}. \\

\noindent \textit{Entropy production for continuous degrees of freedom.} The results above were obtained for Markov jump processes within a discrete state space. However, the decomposition of the rate of change of entropy in Eq.~(\ref{eq:time_der_entr}) into internal and external contributions can be readily generalised to Markovian processes with continuous degrees of freedom, for example a spatial coordinate. For simplicity we will restrict ourselves to processes in one dimension but as far as the following discussion is concerned, generalising to higher dimensions is a matter of replacing spatial derivatives and integrals over the spatial coordinate with their higher dimensional counterparts. The dynamics of such a process with probability density $P(x,t)$ to find it at $x$ at time $t$ are captured by a Fokker-Planck equation of the form $\dot{P}(x,t) = -\partial_x j(x,t)$, with $j$ the probability current, augmented by an initial condition $P(x,0)$. Starting from the Gibbs-Shannon's entropy for a continuous random variable $S(t) = -\int \dint{x}  P(x,t) \ln (P(x,t)/P_0)$ with some arbitrary density scale $P_0$ for dimensional consistency, we differentiate with respect to time and substitute $-\partial_x j(x,t)$ for $\dot{P}(x,t)$ to obtain
\begin{equation}
\label{eq:FP_continuous_ep}
    \dot{S}(t) = -\int \dint{x} \dot{P}(x,t) \ln \left(\frac{P(x,t)}{P_0}\right) = -\int \dint{x} \frac{(\partial_x P(x,t)) j(x,t)}{P(x,t)} ~,
\end{equation}
where the second equality follows upon integration by parts using $\int \dint{x} \dot{P}(x,t)=0$ by normalisation. For the paradigmatic case of an overdamped colloidal particle, which will be discussed in more detail below (Secs.~\ref{sec:DDparticle} --
\ref{sec:Bpartring}), the probability current is given by $j(x,t) = -D \partial_x P(x,t) + \mu F(x,t) P(x,t)$ with local, time-dependent force $F(x,t)$. We can then decompose the entropy production $\dot{S}(t) = \entropyProduction(t) + \entropyFlow(t)$ into internal and external contributions as
\begin{equation}
\label{eq:continuous_entropies_decomp_prod}
    \entropyProduction(t) = \int \dint{x} \frac{j(x,t)^2}{D P(x,t)} \geq 0
\end{equation}
and
\begin{equation}
\label{eq:continuous_entropies_decomp_flow}
    \entropyFlow(t) = - \int \dint{x} \frac{\mu}{D} F(x,t) j(x,t) ~,
\end{equation}
respectively. The Kullback-Leibler divergence between the densities of forward and time-reversed paths can be calculated as outlined above for discrete state systems, thus producing an alternative expression for the internal entropy production in the form 
\begin{align}
 \entropyProduction(t) &= \lim_{\Delta t \to 0}h^R(t,\Delta t) - h(t,\Delta t) \nonumber \\ &= \lim_{\tau \to 0} \frac{1}{2\tau} \int \dint{x} \dint{x'} (P(x,t) W(x\to x',\tau)  - P(x',t) W(x'\to x,\tau) ) \ln \frac{P(x,t) W(x\to x',\tau)  }{P(x',t) W(x'\to x,\tau) } \label{eq:continuum_en_prod} ~.
\end{align}
Here we have introduced the propagator $W(x' \to x,\tau)$, the probability density that a system observed in state $x'$ will be found at $x$ time $\tau$ later. In general, here and above, the density $W(x \to x',\tau)$ depends on the absolute time $t$, which we have omitted here for better readability. 
The corresponding expression for the entropy flow is obtained by substituting (\ref{eq:continuum_en_prod}) into the balance equation $\entropyFlow(t) = \dot{S}(t) - \entropyProduction(t)$, whence
\begin{equation}
\label{eq:continuum_en_flow}
    \entropyFlow(t) =
    - \lim_{\tau \to 0} \frac{1}{2\tau} \int \dint{x} \dint{x'} (P(x,t) W(x \to x',\tau)  - P(x',t) W(x'\to x,\tau) ) \ln \frac{W(x'\to x,\tau) }{W(x\to x',\tau)} ~.
\end{equation}
Since $\lim_{\tau \to 0} W(x \to x',\tau) = \delta(x-x')$ \cite{Wissel1979Jun} and $P(x,t) \delta(x-x') = P(x',t) \delta(x'-x)$ the factor in front of the logarithm in (\ref{eq:continuum_en_prod}) and (\ref{eq:continuum_en_flow}) vanishes in the limit of small $\tau$, $\lim_{\tau \to 0} P(x,t) W(x \to x';\tau) - P(x',t) W(x' \to x;\tau) = 0$. Together with the prefactor $1/\tau$ this necessitates the use of L'Hôpital's rule
\begin{equation}
\label{eq:hopital_continuum}
    \lim_{\tau \to 0} \frac{1}{\tau} \left( P(x,t) W(x \to x';\tau) - P(x',t) W(x' \to x;\tau) \right) = P(x,t) \dot{W}(x \to x') - P(x',t) \dot{W}(x' \to x)
\end{equation}
where we used the shorthand 
\begin{equation}
\label{eq:def_rate}
    \dot{W}(x \to x') := \lim_{\tau \to 0} 
    \frac{\plaind}{\plaind \tau} W(x \to x';\tau) ~,
\end{equation}
which is generally given by the Fokker-Planck equation of the process, so that 
\begin{equation}
    \dot{P}(x,t) = \int \dint{x'} P(x',t) \dot{W}(x'\to x)
    \label{eq:FPE_general} 
    \ .
\end{equation}
In the continuum processes considered below, in particular Sec.~\ref{sec:Bpartring}, \ref{sec:RnTring} and \ref{sec:switchdiff}, $\dot{W}(x \to x')$ is a kernel in the form of Dirac $\delta$-functions and derivatives thereof, acting under the integral as the adjoint Fokker-Planck operator on $P(x,t)$. With \Eref{hopital_continuum} the internal entropy production of a continuous process (\ref{eq:continuum_en_prod}) may conveniently be written as
\begin{subequations}
\elabel{Si_cont_all}
\begin{align}
    \entropyProduction(t) &= \frac{1}{2} \int \dint{x} \dint{x'} \left( P(x,t) \dot{W}(x \to x') - P(x',t) \dot{W}(x' \to x) \right) \times \lim_{\tau \to 0} \ln \left( \frac{P(x,t) W(x \to x';\tau)}{P(x',t) W(x' \to x;\tau)} \right) \label{eq:Si_cont_easy} \\
    &= \int \dint{x} \dint{x'} P(x,t) \dot{W}(x \to x') \times \lim_{\tau \to 0} \ln \left( \frac{P(x,t) W(x \to x';\tau)}{P(x',t) W(x' \to x;\tau)} \right) \\
    &= \int \dint{x} \dint{x'} \left( P(x,t) \dot{W}(x \to x') - P(x',t) \dot{W}(x' \to x) \right) \times \lim_{\tau \to 0} \ln \left( \frac{P(x,t) W(x \to x'; \tau)}{W_0 P_0} \right) 
\end{align}
\end{subequations}
with suitable constants $W_0$ and $P_0$. Correspondingly, the (external) entropy flow (\ref{eq:continuum_en_flow}) is
\begin{subequations}
\elabel{Se_cont_all}
\begin{align}
    \entropyFlow(t) &= -\frac{1}{2} \int \dint{x} \dint{x'} \left( P(x,t) \dot{W}(x \to x') - P(x',t) \dot{W}(x' \to x) \right) \times \lim_{\tau \to 0} \ln \left( \frac{ W(x \to x';\tau)}{ W(x' \to x;\tau)} \right) \label{eq:Se_cont_easy} \\
    &= -\int \dint{x} \dint{x'} P(x,t) \dot{W}(x \to x') \times \lim_{\tau \to 0} \ln \left( \frac{ W(x \to x';\tau)}{ W(x' \to x;\tau)} \right) \\
    &= -\int \dint{x} \dint{x'} \left( P(x,t) \dot{W}(x \to x') - P(x',t) \dot{W}(x' \to x) \right) \times \lim_{\tau \to 0} \ln \left( \frac{ W(x \to x'; \tau)}{W_0} \right) \,.
\end{align}
\end{subequations}
All of these expressions assume that the limits of the logarithms exist. Naively replacing them by $\ln(\delta(x-x')/\delta(x'-x))$ produces a meaningless expression with a Dirac $\delta$-function in the denominator. \Erefs{Si_cont_all} and \eref{Se_cont_all} are identically obtained in the same manner as \Erefs{external_en_prod_and_internal_en_prod} with the
master \Eref{markov_master_eq} replaced by the Fokker-Planck \Eref{FPE_general}. All of these 
expressions, \Eref{external_en_prod_and_internal_en_prod}, \eref{Si_cont_all} and \eref{Se_cont_all}, may thus be seen as Gaspard's \cite{gaspard_time-reversed_2004} framework. \\

\noindent \textit{Langevin description and stochastic entropy. \label{S:fluctuating_entropy}} We have seen 
in \Erefs{continuous_entropies_decomp_prod} and \eref{continuous_entropies_decomp_flow} 
how the notion of entropy production can be extended to continuous degrees of freedom by means of a Fokker-Planck description of the stochastic dynamics. The Fokker-Plank equation is a deterministic equation for the probability density and thus provides a description at the level of ensembles, rather than single fluctuating trajectories. A complementary description can be provided by means of a Langevin equation of motion, which is instead a stochastic differential equation for the continuous degree of freedom \cite{Pavliotis2014}. The presence of an explicit noise term, which usually represents faster degrees of freedom or fluctuations induced by the contact with a heat reservoir, allows for a clearer thermodynamic interpretation. A paradigmatic example is that of the overdamped colloidal particle mentioned above, whose dynamics are described by
\begin{equation}
\label{eq:colloidal_part}
    \dot{x}(t) = \mu F(x,t) + \zeta(t)
\end{equation}
with $\mu$ a mobility, $F(x,t)$ a generic force and $\zeta(t)$ a white noise term with covariance $\langle \zeta(t) \zeta(t') \rangle = 2D \delta(t-t')$. For one-dimensional motion on the real line, the force $F(x,t)$ can always be written as the gradient of a potential $V(x,t)$, namely $F(x,t) = - \partial_x V(x,t)$, so that it is conservative. For time-independent, stable potentials, $V(x,t)=V(x)$, this leads at long times to an equilibrium steady-state. This property does not hold in higher dimensions and for different boundary conditions (e.g.\ periodic), in which case the force $\Fvec(\xvec,t)$ need not have a corresponding potential $V(\xvec,t)$ for which $\Fvec(\xvec,t) = - \nabla V(\xvec,t)$ \cite{wang_2015_landscape_flux}.

The concept of entropy is traditionally introduced at the level of ensembles. However, due to its r{\^o}le in fluctuation theorems \cite{seifert_stochastic_2012,lebowitz_spohn_1999}, a consistent definition at the level of single trajectories is required. 
This can be constructed along the lines of \cite{Seifert:2005} by positing the trajectory-dependent entropy $S(x^*(t),t)$ where $x^*(t)$ is a random trajectory as given by Eq.~(\ref{eq:colloidal_part}) and
\begin{equation}
\label{eq:stoch_entropy}
    S(x,t) = - \ln (P(x,t)/P_0) ~.
\end{equation}
Here $P(x,t)$ denotes the probability density of finding a particle at position $x$ at time $t$ as introduced above and $P_0$ is a scale as used above to maintain dimensional consistency. Given that $x^*(t)$ is a random variable, so is $S(x^*(t),t)$, which may be regarded as an instantaneous entropy. Taking the total derivative with respect to $t$ produces
\begin{align}
\label{eq:tder_fluctuating_en}
    \frac{\plaind}{\plaind t} S(x^*(t),t) &= - \left. \frac{\partial_t P(x,t)}{P(x,t)} \right|_{x=x^*(t)} - \left. \frac{\partial_x P(x,t)}{P(x,t)} \right|_{x=x^*(t)} \circ \dot{x}^*(t) \nonumber \\
    &= - \left. \frac{\partial_t P(x,t)}{P(x,t)} \right|_{x=x^*(t)} + \frac{j(x^*(t),t)}{DP(x^*(t),t)} \circ \dot{x}^*(t) - \frac{\mu}{D} F(x^*(t),t) \circ \dot{x}^*(t)
\end{align}
where we have used the processes' Fokker-Planck equation $\partial_t P(x,t) = - \partial_x j(x,t)$ with $j(x,t) = \mu F(x,t) P(x,t)-D \partial_x P(x,t)$. The total time derivative has been taken as a conventional derivative implying the Stratonovich convention indicated by $\circ$, which will become relevant below. The term in (\ref{eq:tder_fluctuating_en}) containing $\partial_t P(x,t)$ accounts for changes in the probability density due to its temporal evolution, such as relaxation to a steady state, and any time-dependent driving protocol. The product $F(x^*(t),t) \circ \dot{x}^*(t)$ can be interpreted as a power expended by the force and in the absence of an internal energy of the particle, dissipated in the medium. With Einstein's relation defining the temperature of $T=D/\mu$ of the medium, the last term may be written as 
\begin{equation}\elabel{def_Sm}
\dot{S}_m(t) = \frac{F(x^*(t),t) \circ \dot{x}^*(t)}{T}    
\end{equation}
and thus interpreted as the entropy change in the medium.
Together with the entropy change of the particle, this gives the total entropy change of particle and medium,
\begin{equation}
\label{eq:tot_inst_en}
    \dot{S}_{\rm tot} (t) = \frac{\plaind}{\plaind t} S(x^*(t),t) + \dot{S}_m(t) = - \left. \frac{\partial_t P(x,t)}{P(x,t)} \right|_{x=x^*(t)} + \frac{j(x^*(t),t)}{DP(x^*(t),t)} \circ \dot{x}^*(t)\,,
\end{equation}
which is a random variable, as it depends on the position $x^*(t)$. It also draws on $P(x,t)$ and $j(x,t)$ which are properties of the ensemble. To make the connection to the entropies constructed above we need to take an ensemble average of the instantaneous $\dot{S}_{\rm tot}(t)$. To do so, we need an interpretation of the last term of (\ref{eq:tot_inst_en}), where the noise $\zeta(t)$ of $\dot{x}^*(t)$, Eq.~(\ref{eq:colloidal_part}), multiplies $j(x^*(t),t)/P(x^*(t),t)$. Equivalently, we need the joint density $P(x,\dot{x};t)$ of position $x$ and velocity $\dot{x}$ at time $t$. In the spirit of Ito this density trivially factorises into a normally distributed $\dot{x}-\mu F(x,t)$ and $P(x,t)$ as the increment $\dot{x} dt$ on the basis of (\ref{eq:colloidal_part}) depends only on the particle's current position $x(t)$.
However, this is not so in the Stratonovich interpretation of $P(x,\dot{x};t)$, as here the increment depends equally on $x(t)$ and $x(t+dt)$ \cite{seifert_stochastic_2012,SeifertLecturenotes:2008,PietzonkaSeifert:2017}. Taking the ensemble average of $\dot{S}_{\rm tot}$ thus produces
\begin{align}
    \langle \dot{S}_{\rm tot}(t)\rangle &= \int \dint{x^*} \dint{\dot{x}^*} \dot{S}_{\rm tot}(t) P(x^*,\dot{x}^*;t) \nonumber \\
    &= - \int \dint{x^*} \frac{\partial_t P(x^*,t)}{P(x^*,t)} \int \dint{\dot{x}^*} P(x^*,\dot{x}^*;t) + \int \dint{x^*} \dint{\dot{x}^*} \frac{j(x^*,t)}{D P(x^*,t)} \dot{x}^* P(x^*,\dot{x}^*;t)\,,
\end{align}
where $x^*$ and $\dot{x}^*$ are now dummy variables. The first term on the right hand side vanishes, because $P(x^*,t) = \int \dint{\dot{x}^*} P(x^*,\dot{x}^*;t)$ is the marginal of $P(x^*,\dot{x}^*;t)$ and $\int \dint{x^*} \partial_t P(x^*,t)=0$ by normalisation. The integral over $\dot{x}^*$ in the second term produces the expected particle velocity conditional to its position,
\begin{equation}
\label{eq:conditiona_velocity_pos}
    \langle \dot{x}^* | x^*,t \rangle = \int \dint{\dot{x}^*} \dot{x}^* \frac{P(x^*,\dot{x}^*;t)}{P(x^*,t)}
\end{equation}
in the Stratonovich sense, where it gives rise to the current \cite{Seifert:2005}, $\langle \dot{x}^* | x^*,t \rangle = j(x^*,t)/P(x^*,t)$, so that
\begin{equation}
\label{eq:tot_en_prod}
    \langle \dot{S}_{\rm tot} (t) \rangle = \int \dint{x^*} \frac{j^2(x^*,t)}{DP(x^*,t)} \geq 0 ~,
\end{equation}
which vanishes only in the absence of any probability current, \ie in thermodynamic equilibrium. In the Ito sense, the conditional expectation (\ref{eq:conditiona_velocity_pos}) would have instead given rise to the ensemble-independent drift, $\langle \dot{x}^* | x^*,t \rangle = \mu F(x^*,t)$. Comparing to Eq.~(\ref{eq:continuous_entropies_decomp_prod}), the expectation $\langle \dot{S}_{\rm tot}(t) \rangle$ turns out to be the internal entropy production $\entropyProduction(t)$, so that $\dot{S}_{\rm tot}(t)$ of Eq.~(\ref{eq:tot_inst_en}) may be regarded as its instantaneous counterpart. \\

\noindent \textit{Path integral methods.} An interesting aspect of working with the Langevin description is the possibility of casting probability densities $p([x];t)$ for paths $x(t')$ with $t' \in [0,t]$ into path integrals, for example in the Onsager-Machlup formalism \cite{onsager_fluctuations_1953,tauber_2014}. For the colloidal particle introduced in (\ref{eq:colloidal_part}), it gives $p([x];t) = \mathcal{N} \exp{-\mathcal{A}([x];t)}$ with the action functional
\begin{equation}
\label{eq:action_F}
    \mathcal{A}([x];t) = \int_0^t \dint{t'} \frac{(\dot{x}(t')-\mu F(x(t'),t'))^{\circ2}}{4D}
    - \frac{\mu}{2} \int_0^t \dint{t'} \partial_x F(x(t'),t')
\end{equation}
in the Stratonovich discretisation, which differs from the Ito form only by the second term \cite[Sec. 4.5]{tauber_2014}, which is the Jacobian of the transform 
of the noise $\zeta(t)$ to $x(t)$, \Eref{colloidal_part}. 
The Stratonovich form is needed so that the action does not give 
preference to a particular time direction \cite{Cugliandolo_2017}. This choice plays a 
r{\^o}le in every product of white noise, as is implicit to $\dot{x}$, and a random variable. We therefore indicate the choice
by a $\circ$ also in powers, reminding us that $F(x(t'),t')$ should be read as $F((x(t')+x(t'+\Deltat))/2, t'+\Deltat)$
and $\dot{x}(t')$ as $(x(t'+\Deltat)-x(t'))/2$ with discretisation
time step $\Deltat$.
Evaluating the action for the reversed path $x^R(t') = x(t-t')$ then gives
\begin{align}
    \mathcal{A}([x^R ];t) &= \int_0^t \dint{t'} \frac{(\dot{x}^R(t')-\mu F(x^R(t'),t'))^{\circ2}}{4D} 
    - \frac{\mu}{2} \int_0^t \dint{t'} \partial_x F(x^R(t'),t')
    \\
    &= \int_0^t \dint{t'} \frac{(\dot{x}(t') + \mu F(x(t'),t-t'))^{\circ2}}{4D} 
    - \frac{\mu}{2} \int_0^t \dint{t'} \partial_x F(x(t'),t-t')
    ~.
\end{align}
If the force is even under time reversal, $F(x,t') = F(x,t-t')$, 
in particular when it is independent of time,
the path probability density obeys
\begin{equation}
\label{eq:OM_path_log}
    \ln \frac{p([x];t)}{p([x^R];t)} = \int_0^t \dint{t'}   \frac{F(x(t'),t')\circ\dot{x}(t')}{T} = S_m(t) ~,
\end{equation}
with random variables multiplied with Stratonovich convention.
With \Eref{def_Sm},
the integral in \Eref{OM_path_log} 
can be identified as the entropy  of the medium. When the driving is time-independent and the system's probability distribution eventually becomes stationary, such that $\lim_{t \to \infty} \langle \dot{S}(x^*,t) \rangle =0$, Eq.~(\ref{eq:stoch_entropy}), the only contribution to the total entropy change is due to change of entropy in the medium, Eq.~(\ref{eq:tot_inst_en}).
Assuming that the system is ergodic,
we have the equivalence $\lim_{t \to \infty} S_m(t) /t = \lim_{t \to \infty} \langle \dot{S}_{\rm tot}(t) \rangle$,
where $\langle\bullet\rangle$ denotes an
ensemble average.
Using Eqs.~(\ref{eq:continuous_entropies_decomp_prod}) and (\ref{eq:tot_en_prod})
gives $\lim_{t \to \infty} S_m(t)/t = \lim_{t \to \infty}  \entropyProduction(t)$.
Equation (\ref{eq:OM_path_log}) can therefore be used directly to compute the steady-state internal entropy production rate. The equivalence between the long-time limit $t \to \infty$ and the ensemble average holds only for ergodic systems, whose unique steady-state does not depend on the specific initialisation $x(0)$.
This connection between stochastic thermodynamics and field theory has stimulated a number of works aimed at characterising the non-equilibrium features of continuum models of active matter \cite{nardini_entropy_2017,fodor_how_2016}.

\section{Systems \label{S:systems}}

In this section we calculate the entropy production rate on the basis of Gaspard's framework \cite{gaspard_time-reversed_2004}, Eqs.~(\ref{eq:external_en_prod_and_internal_en_prod}), (\ref{eq:continuum_en_prod}) and (\ref{eq:continuum_en_flow}), for different
particle systems. We cover the systems listed in  Tab.~\ref{tab:my_label}, with both discrete and continuous
states and with one or multiple particles.

\begin{table}[h]
    \caption{List of particle systems for which we have calculated their entropy production
    $\entropyProduction(t)$.}
    \label{tab:my_label}
    \centering
    \begin{tabular}{l l c c c c }
        & System 
        & $\entropyProduction(t)$ \\ 
        \hline
        \ref{sec:sec2state} & Two-state Markov process 
        & \eref{si2state} \\ 
        \ref{sec:sec3state} & Three-state Markov process 
        & \eref{si3state} \\ 
        \ref{sec:N1d}& Random walk on a complete graph 
        & \eref{siRWd1}, \eref{siRWd2} \\ 
        \ref{sec:Ndist} & $N$ independent, distinguishable Markov processes 
        &(\ref{eq:Si_Ndist}) \\
        \ref{sec:N2p} & $N$ independent, indistinguishable two-state Markov processes 
        & \eref{N2indist} \\
        \ref{sec:Nindepd} & $N$ independent, indistinguishable $d$-state processes 
        &(\ref{eq:N_indist_dstate})\\
        \ref{sec:CTRW1d} & Random Walk on a lattice 
        &\eref{ent_prod_CTRW_expr}\\
        \ref{sec:RWring} & Random Walk on a ring lattice 
        &\eref{Si_ring_lat}, (\ref{eq:steady_Si_ring_lattice})\\
        \ref{sec:DDparticle} & Driven Brownian particle  
        &\eref{siDD}\\
        \ref{sec:OU}& Driven Brownian particle  in a harmonic potential 
        &\eref{OUsiG}\\
        \ref{sec:Bpartring} & Driven Brownian particle  on a ring with potential 
        &(\ref{eq:Si_ring_BP_j})\\
        \ref{sec:RnTring} & Run-and-tumble motion with diffusion on a ring 
        &(\ref{eq:rnt_steady_si})\\
        \ref{sec:switchdiff} & Switching diffusion process on a ring &(\ref{eq:en_prod_switch_diff})
    \end{tabular}
\end{table}

\subsection{Two-state Markov process}
\label{sec:sec2state}

\begin{figure}[h]
\centering
\includegraphics[width=0.3\textwidth]{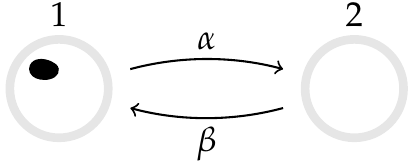}
\caption{\flabel{F1}
Two-state Markov chain in continuous time. The black blob indicates the current state of the system. Independently of the choice of $\alpha$ and $\beta$, this processes settles into an equilibrium steady-state at long times (in the absence of an external time-dependent diving).}
\end{figure}

Consider a particle that hops between two states, $1$ and $2$,  with 
transition rates $\dot{W}(1\to 2)=\alpha$ and $\dot{W}(2\to 1)=\beta$, see \fref{F1} \cite{Esposito:2010,Lesne:2014}, 
and using the notation in \Eref{def_rate} for
discrete states. 
The rate-matrix (see \Eref{W_expanded}) may thus be 
\begin{equation}
\elabel{W2}
     w= \begin{pmatrix}
    -\alpha & \alpha \\
    \beta & -\beta\\
    \end{pmatrix} \,,
\end{equation}
with  $\Pvec(t)=(P_1(t),P_2(t))$ the probability  of the particle to be in state $1$ or $2$ respectively
as a function of time. By normalisation, $P_1(t)+P_2(t)=1$,
with probabilistic initial condition  $\Pvec(0)=(p, 1-p)$.
Solving the master equation in \Eref{markov_master_eq} yields
\begin{equation}
\label{eq:sol_ME_2state}
\Pvec(t) = (P_1(t),P_2(t))
=
\frac{1}{\alpha+\beta}
\begin{pmatrix}
    {\beta}+r\,e^{-(\alpha+\beta) t}, 
    {\alpha}-r\,e^{-(\alpha+\beta) t}
    \end{pmatrix} \,,
\end{equation}
with $r=\alpha p-\beta (1-p)$,
corresponding to an exponentially decaying probability current
\begin{equation}
    P_1(t) \alpha - P_2(t) \beta = r\,e^{-(\alpha + \beta)t} ~.
\end{equation}
The internal entropy production  \eref{internal_en_prod} is then 
\begin{equation}
     \entropyProduction (t) =
     [P_1(t)\alpha-P_2(t)\beta]\ln\left[
     \frac{P_1(t)\alpha}{P_2(t)\beta}
     \right]
     =r \exp{-(\alpha+\beta)t}\ln\left[{\frac{1+ \frac{r}{\beta}\,e^{-(\alpha+\beta)t}}{1 -\frac{r}{\alpha}\,e^{-(\alpha+\beta)t}}}\right] \,,
     \elabel{si2state}
\end{equation}
and the entropy flow  \eref{external_en_prod},
\begin{equation}
    \entropyFlow(t) =-r\,\exp{-(\alpha+\beta)t}\,\ln{\left(\frac{\alpha}{\beta}\right)} \,.
     \elabel{se2state}
\end{equation}
At stationarity, $\entropyProduction=\entropyFlow=0$ and therefore
 the two-state Markov process reaches equilibrium. 
In this example, the topology of the transition network does not allow
a sustained current between states, which inevitably leads to 
equilibrium in the steady state and, therefore, there is production of
entropy only due to the relaxation of the system
from the initial state.

\subsection{Three-state Markov process}
\label{sec:sec3state}

\begin{figure}[h]
\centering
\includegraphics[width=0.3\textwidth]{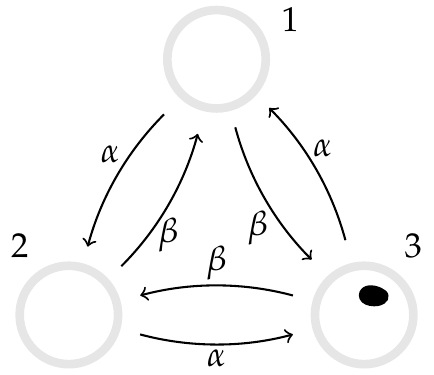}
\caption{\flabel{F2} Three-state Markov chain in continuous time. The black blob indicates the current state of the system. Symmetry under cyclic permutation is introduced by imposing identical transition rates $\alpha$ and $\beta$ for counter-clockwise and clockwise transition, respectively.}
\end{figure}   

We extend the system in \sref{sec2state} to three states,
$1$, $2$ and $3$, with transition rates 
$\dot{W}(1\to 2)=\alpha$, 
$\dot{W}(2\to 3)=\alpha$, 
$\dot{W}(3\to 1)=\alpha$, 
$\dot{W}(2\to 1)=\beta$,  
$\dot{W}(3\to 2)=\beta$, and 
$\dot{W}(1\to 3)=\beta$, see \fref{F2}, and using the notation \Eref{def_rate} for
discrete states. The
 rate matrix (see \Eref{W_expanded}) is then
\begin{equation}
     w = \begin{pmatrix}
    -(\alpha+\beta) & \alpha & \beta \\
    \beta& -(\alpha+\beta) & \alpha \\
    \alpha & \beta & -(\alpha+\beta)
    \end{pmatrix} \,.
    \elabel{transmat3}
\end{equation}
Assuming the initial condition  $\Pvec(0)=(1, 0, 0)$,
the probabilities of states 1, 2 and 3 
respectively, evolve according to \Eref{markov_master_eq}, which has solution
\begin{subequations}
\elabel{distr3state}
\begin{align}
    P_{1}(t)= &\frac{1}{3}\left(1+ 2 \exp{-3\phi t} \cos{(\sqrt{3}\psi t)}\right) \,,\\
P_{2}(t)= &\frac{1}{3}\left(1- 2 \exp{-3\phi t} \cos{(\sqrt{3}\psi t -\pi/3)}\right) \,,\\
P_{3}(t)= &\frac{1}{3}\left(1- 2 \exp{-3\phi t} \cos{(\sqrt{3}\psi t +\pi/3)}\right)\,,
\end{align}
\end{subequations}
with $\phi=(\alpha+\beta)/2$ and $\psi=(\alpha-\beta)/2$.

The entropy production \eref{internal_en_prod} is then, using \eref{distr3state}, 
\begin{multline}
\entropyProduction (t) =
 \big(P_{1}(t)\alpha-P_{2}(t)\beta\big)
\ln{\left({\frac{P_{1}(t)\alpha}{P_{2}(t)\beta}}\right)}\\
+\big(P_{2}(t)\alpha-P_{3}(t)\beta\big)
\ln{\left({\frac{P_{2}(t)\alpha}{P_{3}(t)\beta}}\right)}
+\big(P_{3}(t)\alpha-P_{1}(t)\beta\big)
\ln{\left({\frac{P_{3}(t)\alpha}{P_{1}(t)\beta}}\right)}\,,
     \elabel{si3state}
\end{multline}
and the entropy flow \eref{external_en_prod},
\begin{equation}
    \entropyFlow(t)=-(\alpha-\beta)\ln{\left(\frac{\alpha}{\beta}\right)} \,,
     \elabel{se3state}
\end{equation}
which is constant throughout.
At stationarity, the system is uniformly distributed and, if
$\alpha\neq\beta$, the entropy
production and flow satisfy $\entropyProduction=-\entropyFlow\neq0$.
If $\alpha\neq\beta$, the particle has a net drift that sustains a 
probability current $(\alpha-\beta)/3$ in the system, which prevents the system from reaching
equilibrium.


\subsection{Random walk on a complete graph}
\label{sec:N1d}

\begin{figure}[h]
\centering
\includegraphics[width=0.4\textwidth]{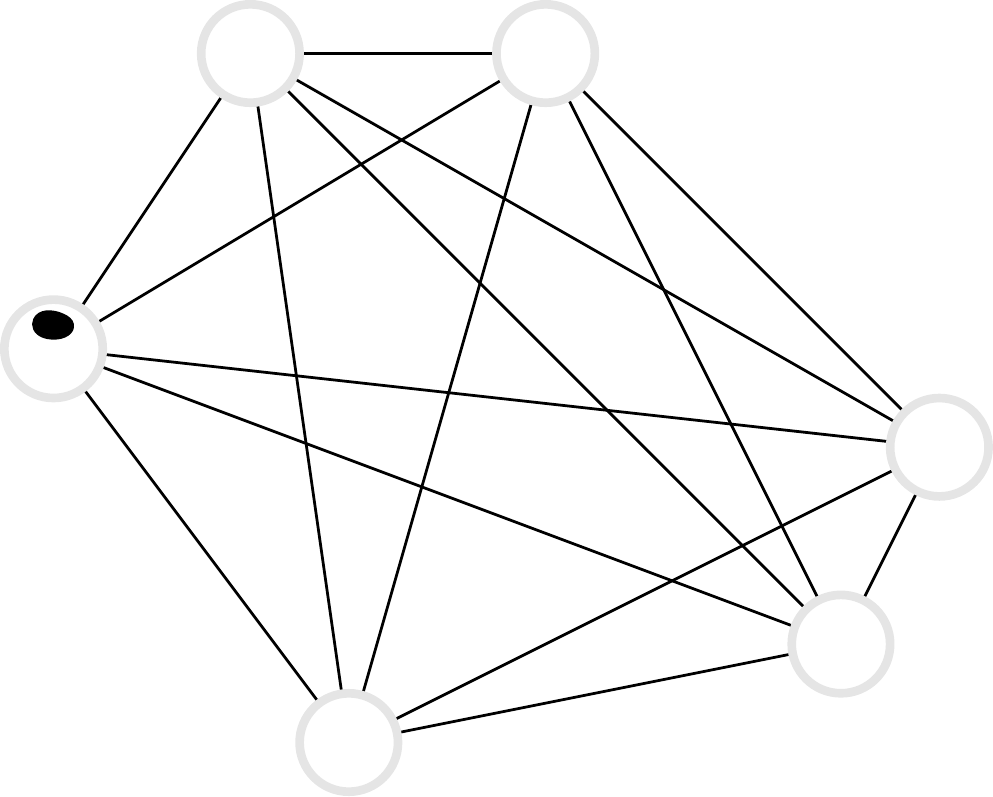}
\caption{\flabel{F3} Random walk on a complete graph of $d$ nodes (here shown for $d=6$). The black blob indicates the current state of the system. For uniform transition rates, the symmetry under node relabelling leads to an equilibrium, homogeneous steady-state with $P_j=1/d$ for all $j$.}
\end{figure}   

\newcommand{\Nstates}{\Omega}
Consider a random walker on a complete graph with $ d $ nodes, 
where each node is connected to all other nodes, 
and the walker jumps from
node $j \in \{1,2,\ldots,d\}$ to node $k\in \{ 1,2,\ldots,d\}$, $k \neq j$, with rate $w_{jk}$, see \fref{F3}. 
These are the off-diagonal elements of the corresponding Markov matrix whose diagonal elements are $w_{jj} = -\sum_{i=1,i\neq j}^d w_{ji}$.
The probability vector 
$\mathbf{P}(t)=(P_1(t), P_2(t),\ldots,P_d(t))$
has components $P_j(t)$ that are the probability that the system is in state $j$ at time $t$.
The general case of arbitrary transition rates is impossible to discuss exhaustively. In the uniform case, $w_{jk} = \alpha $, the Markov matrix has only two distinct eigenvalues, namely 
eigenvalue $ \alpha  d $ with degeneracy $ d -1$ and eigenvalue $  0$ with degeneracy $1$. 
Assuming an arbitrary initial condition $\Pvec(0)$,
the probability distribution at a later time $t$ is 
\begin{equation}
    P_j(t)=  \frac{1}{ d } + e^{- d  \alpha t} \left(P_j(0) - \frac{1}{ d  } \right) \,.
\end{equation}
The steady state, which is associated with the vanishing eigenvalue,
is the uniform distribution $\lim_{t \to \infty} P_j(t)=1/d$ for all $j \in \{1,2,\ldots,d\}$.
The entropy production \eref{internal_en_prod} of the initial state relaxing to the uniform state is
\begin{equation}
    \entropyProduction(t) = \frac{1}{2}\alpha  e^{- d  \alpha t} \sum_{j,k}  
    \left( P_j(0)-P_k(0) \right) \ln \left( 
    \frac{1 + e^{- d  \alpha t} \left( P_j(0)  d  - 1 \right)}{1 + e^{- d  \alpha t} \left( P_k(0)  d  - 1  \right)} \right) \,,
    \elabel{siRWd1}
\end{equation}
and the entropy flow \eref{external_en_prod} is $\entropyFlow=0$ throughout.
If the walker is initially located on node $k$, so that $P_j(0)=\delta_{j,k}$, the entropy production simplifies to
\begin{equation}
    \entropyProduction(t) = ( d  -1)\alpha  e^{- d  \alpha t} \ln(1 + \frac{ d   e^{- d  \alpha t} }{1-e^{- d  \alpha t}}) \,.
    \elabel{siRWd2}
\end{equation}
We can see that the system reaches equilibrium at stationarity, 
since $\lim_{t\to\infty}\entropyProduction(t)=\entropyFlow(t)=0$.
At long times ($ d   e^{- d  \alpha t} \ll 1$), the asymptotic behaviour of $\entropyProduction$ is
\begin{equation}
    \entropyProduction(t) = d  ( d  -1)\alpha  e^{-2 d  \alpha t} + \mathcal{O}(e^{-3d\alpha t})~,
\end{equation}
by expanding the logarithm in the small exponential.

\subsection{$N$ independent, distinguishable Markov processes}
\label{sec:Ndist}

\begin{figure}[h]
\centering
\includegraphics[width=0.75\textwidth]{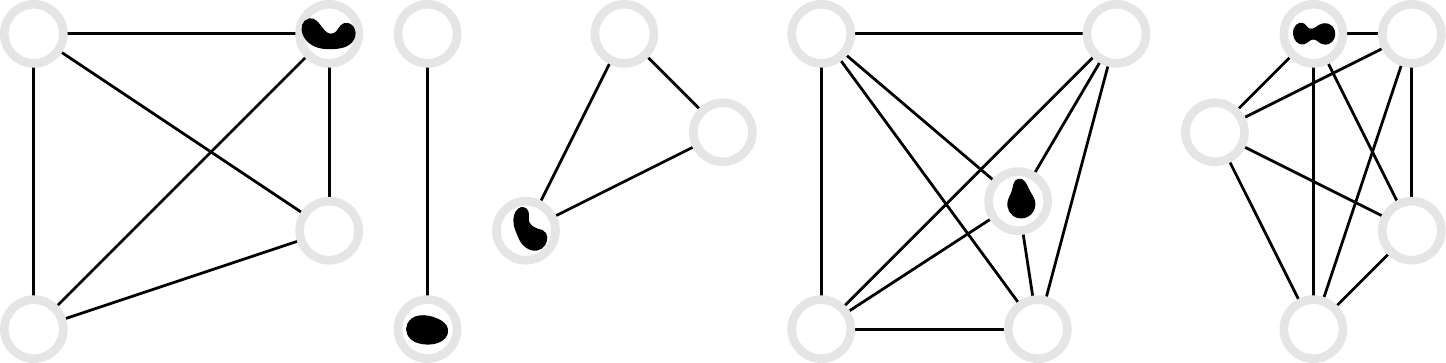}
\caption{\flabel{F4} 
Example of $N=5$ non-interacting,
distinguishable processes with
$d_1=4$, $d_2=2$, $d_3=3$, $d_4=5$ and $d_5=5$. The black blobs indicate the current state of each sub-system.
}
\end{figure}   

In the following we consider $N$ non-interacting, distinguishable 
particles undergoing
Markovian dynamics on a discrete state space. 
Each of the $N$ particles carries an index $\ell\in \{ 1,2,\ldots,N \}$ 
and is in state $n_{\ell}\in\{1,2,\ldots,d_{\ell}\}$, so that the 
state of the entire system is given by an $N$-particle state $\nvec=(n_1,n_2,\ldots,n_N)$.
Particle distinguishability implies the factorisation of state and transition probabilities into their single-particle contributions, whence the 
joint probability $P_\nvec(t)$ of an $N$-particle state $\nvec$ factorises into a product of single particle probabilities $P_{n_\ell}^{(\ell)}(t)$ of particle $\ell$ to be in state $n_\ell$,
\begin{equation}
    P_\nvec(t) = \prod_{\ell=1}^N P_{n_\ell}^{(\ell)}(t) ~.
\end{equation}
Further, the Poissonian rate $w_{\nvec \mvec}$ from $N$-particle state $\nvec$ to $N$-particle state $\mvec \neq \nvec$ vanishes for all transitions $\nvec \to \mvec$ that differ in more than one component $\ell$, \ie $w_{\nvec \mvec}=0$ unless there exists a single $\ell \in \{1,2,\ldots,N \}$ such that $m_k = n_k$ for all $k \neq \ell$, in which case $w_{\nvec \mvec}=w_{n_\ell m_\ell}^{(\ell)}$, the transition rates of the single particle transition of particle $\ell$.

The entropy production of this $N$-particle system according to Eq.~(\ref{eq:internal_en_prod}),
\begin{equation}
    \label{eq:S_nm}
    \entropyProduction(t) = \frac{1}{2} \sum_{\nvec \mvec} (P_\nvec(t) w_{\nvec \mvec} - P_\mvec(t) w_{\mvec \nvec}) \ln \left( \frac{P_\nvec(t) w_{\nvec \mvec}}{P_\mvec(t) w_{\mvec \nvec}} \right)
\end{equation}
simplifies considerably due to $w_{\nvec \mvec}$, as the sum may be re-written as
\begin{equation}
    \sum_{\nvec \mvec} \ldots w_{\nvec \mvec} \ldots = \sum_\nvec \sum_\ell \sum_{m_\ell} \ldots w_{\nvec \mvec_\ell} \ldots
\end{equation}
with $\mvec_\ell = (n_1,n_2,\ldots,n_{\ell-1},m_\ell,n_{\ell+1},\ldots,n_N)$ so that $w_{\nvec \mvec_\ell} = w_{n_\ell m_\ell}^{(\ell)}$ and 
\begin{equation}\elabel{Si_intermediate}
    \entropyProduction(t) = \frac{1}{2} \sum_\nvec \sum_{\ell=1}^N \sum_{m_\ell} \left\{ \left( \prod_{k=1}^N P^{(k)}_{n_k}(t) \right) w_{n_\ell m_\ell}^{(\ell)} - \left( \prod_{k=1}^N P^{(k)}_{m_k}(t) \right) w_{m_\ell n_\ell}^{(\ell)} \right\}  \ln \left( \frac{\prod_{k=1}^N P^{(k)}_{n_k}(t) w^{(\ell)}_{n_\ell m_\ell}  }{\prod_{k=1}^N P^{(k)}_{m_k}(t) w^{(\ell)}_{m_\ell n_\ell} } \right) ~.
\end{equation}
Since $m_k = n_k$ for any $k \neq \ell$ inside the curly bracket,
we may write
\begin{align}
    \prod_{k=1}^N P^{(k)}_{n_k}(t) =
    P^{(\ell)}_{n_\ell}(t) \prod_{\substack{k=1\\ k\ne\ell}}^N P^{(k)}_{n_k}(t)  
    && \text{ and } &&
    \prod_{k=1}^N P^{(k)}_{m_k}(t) =
    P^{(\ell)}_{m_\ell}(t) \prod_{\substack{k=1\\ k\ne\ell}}^N P^{(k)}_{n_k}(t) \ .
\end{align}
The product $\prod_{k \neq \ell}^N P^{(k)}_{n_k}(t)$ can thus
be taken outside the curly bracket in \Eref{Si_intermediate} 
and be summed over, as well as cancelled in the logarithm. After changing the dummy variables in the remaining summation from $n_\ell$ and $m_\ell$ to $n$ and $m$ respectively, the entropy production is
\begin{equation}
\label{eq:Si_Ndist}
    \entropyProduction(t) = \frac{1}{2} \sum_{\ell = 1}^N \sum_{nm} \Big(P_n^{(\ell)}(t) w_{nm}^{(\ell)} - P_m^{(\ell)}(t) w_{mn}^{(\ell)}\Big) \ln \left( \frac{P_n^{(\ell)}(t) w_{nm}^{(\ell)}}{P_m^{(\ell)}(t) w_{mn}^{(\ell)}} \right) \,,
 \end{equation}
which is the sum of the entropy productions of the single particle $\ell \in\{ 1,2,\ldots,N\}$, Eq.~(\ref{eq:internal_en_prod}), 
irrespective of how each particle is initialised. The same argument applies to $\entropyFlow$, the entropy flow Eq.~(\ref{eq:external_en_prod}). The entropy production and flow obviously simplify to an $N$-fold product of the single particle expressions if $w_{nm}^{(\ell)}$ do not depend on $\ell$ and all particles are initialised by the same $P_n^{\ell}(0)$ independent of $\ell$. This result may equally be found from\textbf{} the dynamical entropy per unit time, Eq.~(\ref{eq:h_discrete}).

\subsection{$N$ independent, indistinguishable two-state Markov processes}
\label{sec:N2p}

\begin{figure}[h]
\centering
\includegraphics[width=0.3\textwidth]{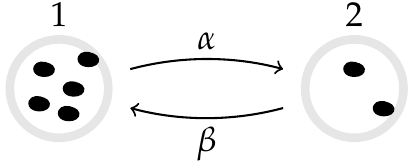}
\caption{\flabel{F5} $N$ independent, indistinguishable two-state Markov processes in continuous time. The black blobs indicate the current state of the single-particle sub-system. Since processes are indistinguishable, states are fully characterised by the occupation number of either state, if the total number of particles is known.}
\end{figure}   

Suppose that $N$ identical, indistinguishable, non-interacting particles  follow the two-state Markov process described in \sref{sec2state}, \fref{F5} \cite{Esposito:2010}.
There are $\Omega = N+1$ distinct states given by the occupation number $n \in \{0,1,\ldots,N\}$ of one of the two states, say state $1$, as the occupation number of the other state follows as $N-n$ given the particle number $N$ is fixed under the dynamics. 
In the following, $P(n,t)$ denotes the probability of finding $n$ particles in state 1 at time $t$. 
The master equation is then
\begin{equation}
\label{eq:N_id_MR}
    \dot{P}(n,t) = -\alpha n P(n,t) + \alpha (n+1) P(n+1,t) -\beta (N-n) P(n,t) + \beta (N-n+1) P(n-1,t) ~.
\end{equation}
The state space and the evolution in it can be thought of as a hopping process on a one-dimensional chain of states with non-uniform rates. Provided $P(n,0)$ initially follows a binomial distribution, $P(n,0)={N \choose n} p^n (1-p)^{N-n}$ with probability $p$ for a particle to be placed in state $1$ initially,
the solution of Eq.~(\ref{eq:N_id_MR}) is easily constructed from the solution $P_1(t)$ in Eq.~(\ref{eq:sol_ME_2state}) of Sec.~\ref{sec:sec2state} via
\begin{equation}
\label{eq:N_id_solution}
    P(n,t) = {N \choose n} P_1^n(t) (1-P_1(t))^{N-n} \ \ \ \text{for} \ \ 0 \leq n \leq N
\end{equation}
with $P_1(0)=p$, as $\dot{P}_1(t) = -\alpha P_1(t) + \beta (1-P_1(t))$, which can be verified by substituting \Eref{N_id_solution} into Eq.~(\ref{eq:N_id_MR}).
Using \Erefs{W2} and (\ref{eq:N_id_solution}) in \eref{internal_en_prod}
the entropy production  reads
\begin{subequations}
\begin{align}
    \entropyProduction(t)
    &= \sum_{n=1}^N [P(n,t) \alpha n - P(n-1,t)\beta(N-n+1)] \ln \left[ \frac{P(n,t)\alpha n}{P(n-1,t)\beta (N-n+1)} \right] \\
    &= N [P_1(t) \alpha - (1-P_1(t))\beta] \ln \left[ \frac{P_1(t) \alpha}{(1-P_1(t))\beta} \right] ~,
    \elabel{N2indist}
\end{align}
\end{subequations}
which is the $N$-fold multiple of the result of the corresponding single particle system, Eq.~\eref{si2state}. This result, Eq.~\eref{N2indist}, depends on the initialisation being commensurable with Eq.~(\ref{eq:N_id_solution}) which otherwise is recovered only asymptotically and only if the stationary distribution is unique. Further, 
the entropy production of $N$ indistinguishable particles being the $N$-fold entropy production of a single particle
does not extend to the external entropy flow, which lacks the simplification of the logarithm and gives
\begin{align}
\label{eq:Se_2state}
    \entropyFlow(t) = - N [\alpha P_1(t) - \beta(1-P_1(t))] \left\{ \ln \left( \frac{\alpha}{\beta} \right) + \sum_{n=0}^{N-1} P_1^n(t) (1-P_1(t))^{N-1-n}  {N-1 \choose n} \ln \left( \frac{n+1}{N-n} \right)\right\} 
\end{align}
thus picking up a correction in the form of the additional sum in the curly bracket that vanishes only at $N=1$ or $P_1(t) = 1/2$, but does not contribute at stationarity because of the overall prefactor $\alpha P_1 -\beta (1-P_1)$ that converges to 0. To make sense of this correction in relation to particle indistinguishability, with the help of \Eref{N_id_solution} we can rewrite the difference between the right hand side of Eq.~(\ref{eq:Se_2state}) and the $N$-fold entropy flow of a single two-state system \eref{se2state} as
\begin{align}
    - &N [\alpha P_1(t) - \beta(1-P_1(t))] \sum_{n=0}^{N-1} P_1^n(t) (1-P_1(t))^{N-1-n}  {N-1 \choose n} \ln \left( \frac{n+1}{N-n} \right)  \nonumber \\
    &= - \sum_{n=0}^{N-1} [\alpha(n+1)P(n+1,t) - \beta(N-n)P(n,t)] \ln \left( \frac{n+1}{N-n} \right) \label{eq:boltz_contrib_se}
\end{align}
which now explicitly involves the net probability current from the occupation number state with $n+1$ particles in state $A$ to that with $n$ particles in state $A$, as well as a the logarithm
\begin{equation}
\label{eq:log_boltzmann_diff}
    \ln \left( \frac{n+1}{N-n} \right) = \ln \left[ {N \choose n} \right] - \ln \left[ {N \choose n+1}  \right] ~.
\end{equation}
Written in terms of the same combinatorial factors appearing in Eq. (\ref{eq:N_id_solution}), the logarithm (\ref{eq:log_boltzmann_diff}) can be interpreted as a difference of microcanonical (Boltzmann) entropies, defined as the logarithm of the degeneracy of the occupation number state if we were to assume that the $N$ particles are distinguishable. 
With the help of the master \Eref{N_id_MR} as well as \Erefs{N_id_solution} and \eref{log_boltzmann_diff}, the term \Eref{boltz_contrib_se} may be rewritten to give
\begin{equation}
      \entropyFlow(t) = - N [\alpha P_1(t) - \beta(1-P_1(t))]  \ln \left( \frac{\alpha}{\beta} \right) 
      - \sum_{n=0}^{N-1} \dot{P}(n,t) \ln \left[{N \choose n} \right]
\end{equation}
This result is further generalised in \Eref{generalised_extra_term_Nid}.

\subsection{$N$ independent, indistinguishable $d$-state processes}
\label{sec:Nindepd}

\begin{figure}[h]
\centering
\includegraphics[width=0.4\textwidth]{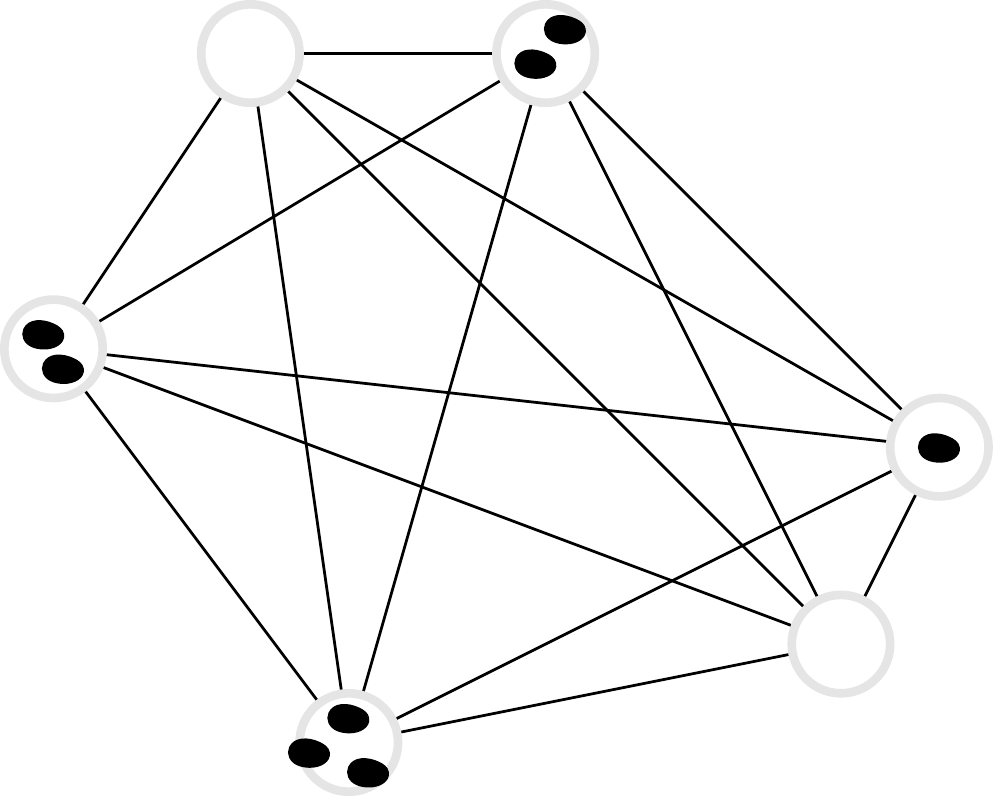}
\caption{\flabel{F6} $N$ independent, indistinguishable $d$-state Markov processes (here shown for $d=6$ and $N=8$) in continuous time. Black blobs indicate the current state of the single-particle sub-systems. Due to indistinguishably, multi-particle states are fully characterised by the occupation number of an arbitrary subset of $d-1$ states, if the total number of particles is known.}
\end{figure}   

We generalise now the results in \sref{N1d} and \sref{N2p}  
to $N$ independent $d$-state Markov processes, see \fref{F6}.
These results represent a special case of those obtained in \cite{herpich_stochastic_2020} when the $N$ processes are non-interacting.
In this section we consider non-interacting, indistinguishable particles hopping on a graph of $d$ nodes with edge-dependent hopping rates
 $w_{jk}$. 
 As in the two-state system in \sref{N2p}, we find that the internal (but not the external) entropy production of the $d$-state system $\entropyProduction$
 is $N$ times the entropy production of the individual processes assuming the initial condition is probabilistically identical for all single-particle sub-systems.
 The entropy productions of a single such process according to \Eref{external_en_prod_and_internal_en_prod} read
\begin{subequations}
\begin{align}
\entropyProduction^{(1)}(t) &= \frac{1}{2} \sum_{jk} [P_j(t)w_{jk} - P_k(t) w_{kj}] \ln \left( \frac{P_j(t) w_{jk}}{P_k(t) w_{kj}} \right) \,, \label{eq:si_single_part}\\
\entropyFlow^{(1)}(t) &= - \frac{1}{2} \sum_{jk} [P_j(t) w_{jk} - P_k(t) w_{kj}] \ln \left( \frac{w_{jk}}{w_{kj}} \right) \,, \label{eq:se_single_part}
\end{align}
\end{subequations}
where $P_j(t)$ is the time-dependent probability of a single-particle process to be in state $j$, \sref{N1d}.
To calculate the entropy production of the $N$ concurrent indistinguishable processes using the occupation number representation,
 we first derive the probability of an occupation number configuration $\nvec = ( n_1, n_2, \ldots, n_d )$, with $\sum_{j=1}^d n_j =N$, which similar to \Eref{N_id_solution} is given by the multinomial distribution
\begin{equation}
\label{eq:multinom_prob}
    P_\nvec(t) = N! \prod_{j=1}^d \frac{P_j^{n_j}(t)}{n_j!} 
\end{equation}
for the probability $P_\nvec(t)$ of the system to be in state $\nvec$ at time $t$
assuming that each particle is subject to the same single-particle distribution $P_j(t)$, $j \in \{ 1,2,\ldots,d\}$ for all $t$, \ie
 in particular assuming that all particles are initialised identically, by placing them all
at the same site or, more generally, by placing them initially according to 
the same distribution $P_j(0)$. 
Given this initialisation, \Eref{multinom_prob} solves \Eref{markov_master_eq}
\begin{equation}\elabel{ME_Nindindd}
    \dot{P}_\nvec(t) = \sum_{\mvec} P_\mvec(t) w_{\mvec\nvec} - P_\nvec(t) w_{\nvec\mvec}
\end{equation}
with the transition rates $w_{\mvec\nvec}$ discussed below.

For non-interacting processes with a unique stationary distribution, \Eref{multinom_prob} is always obeyed in the limit of long times after initialisation, since the single-particle distributions $P_j(t)$ are identical at steady state.
The entropy production Eq.~\eref{internal_en_prod} of the entire system has the same form as Eq.~(\ref{eq:S_nm}) of Sec.~\ref{sec:Ndist} ($N$ independent, \emph{distinguishable} particles) with $w_{\nvec \mvec}$ however now the transition rate between the occupation number state $\nvec = (n_1,n_2,\ldots,n_d)$ with $0 \leq n_k \leq N$ to occupation number state $\mvec = (m_1,m_2,\ldots,m_d)$. The rate $w_{\nvec \mvec}$ vanishes except when $\mvec$ differs from $\nvec$ in exactly two distinct components, say $m_j = n_j-1 \geq 0$ and $m_k = n_k+1 \geq 1$ in which case $w_{\nvec \mvec} = n_j w_{jk}$ with $w_{jk}$ the transition rates of a single particle from $j$ to $k$ as introduced above. For such $\mvec$, the rate obeys $w_{\mvec \nvec} = m_k w_{kj}$ and the probability $P_\mvec(t)$ fulfills 
\begin{equation}
    P_\mvec(t) = P_\nvec(t) \frac{P_k(t) n_j}{P_j(t) m_k} = P_\nvec(t) \frac{P_k(t) w_{\nvec \mvec} w_{kj} }{P_j(t) w_{\mvec \nvec} w_{jk}} ~,
\end{equation}
which simplifies the entropy production Eq.~(\ref{eq:S_nm}) to
\begin{align}
\label{eq:N_ii_d_si}
    \entropyProduction(t) &= \frac{1}{2} \sum_{\nvec \mvec}(P_\nvec(t) w_{\nvec \mvec} - P_\mvec(t) w_{\mvec \nvec} ) \ln \left( \frac{P_\nvec(t) w_{\nvec \mvec}}{P_\mvec(t) w_{\mvec \nvec}} \right) \nonumber \\
    &= \frac{1}{2} \sum_\nvec \sum_{jk} (P_\nvec(t) n_j w_{jk} - P_\mvec(t) m_k w_{kj}) \ln \left( \frac{P_j(t) w_{jk}}{P_k(t) w_{kj}} \right)
\end{align}
where the sum $\sum_\nvec$ runs over all allowed configurations, namely $0 \leq n_j \leq N$ for $j=1,2,\ldots,d$ with $\sum_j n_j = N$ and $\mvec=(n_1,n_2,\ldots,n_j-1,\ldots,n_k+1,\ldots,n_d)$ is derived from $\nvec$ as outlined above. Strictly, $P_\nvec(t)$ has to be defined to vanish for invalid states $\nvec$, so that the first bracket in the summand of Eq.~(\ref{eq:N_ii_d_si}) vanishes in particular when $n_j=0$, in which case $m_j=-1$. To proceed, we introduce the probability
\begin{equation}
    \bar{P}_{\bar{\nvec}_j}(t) = (N-1)! \frac{P_j^{n_j -1}}{(n_j - 1)!} \prod_{i=1, i \neq j}^d \frac{P_i^{n_i}}{n_i!}\ ,
\end{equation}
defined to vanish for $n_j = 0$, so that $P_\nvec(t) n_j = N P_j(t) \bar{P}_{\bar{\nvec}_j}(t)$. The probability $\bar{P}_{\bar{\nvec}_j}(t)$ is that of finding $n_i$ particles at states $i \neq j$ and $n_j -1$ particles at state $j$. It is Eq.~(\ref{eq:multinom_prob}) evaluated in a system with only $N-1$ particles and configuration $\bar{\nvec}_j = (n_1,n_2,\ldots,n_{j-1},n_j -1, n_{j+1},\ldots,n_d) = \bar{\mvec}_k$ a function of $\nvec$. Eq.~(\ref{eq:N_ii_d_si}) may now be rewritten as
\begin{equation}
    \entropyProduction(t) = \frac{N}{2} \sum_{jk} \left\{ \left( \sum_\nvec \bar{P}_{\bar{\nvec}_j}(t) \right) P_j(t) w_{jk} - \left( \sum_\nvec \bar{P}_{\bar{\mvec}_k}(t) \right) P_k(t) w_{kj} \right\} \ln \left( \frac{P_j(t) w_{jk}}{P_{k}(t) w_{kj}} \right)
\end{equation}
where we have used that the arguments of the logarithm are independent of $\nvec$ and $\mvec$. The summation over $\nvec$ gives
\begin{equation}
    \sum_\nvec \bar{P}_{\bar{\nvec}_j}(t) = \sum_\nvec \bar{P}_{\bar{\mvec}_k}(t) = 1
\end{equation}
so that
\begin{equation}
    \entropyProduction(t) = \frac{N}{2} \sum_{jk} (P_j(t) w_{jk} - P_k(t) w_{kj}) \ln \left( \frac{P_j(t) w_{jk}}{P_k(t) w_{kj}} \right) = N \entropyProduction^{(1)}(t)
    \label{eq:N_indist_dstate}
\end{equation}
which is the $N$-fold entropy production of the single particle system $\entropyProduction(t)$, Eq.~(\ref{eq:si_single_part}), or equivalently that of $N$ distinguishable particles, Eq.~(\ref{eq:Si_Ndist}), Sec.~\ref{sec:Ndist}.
As in Sec.~\ref{sec:N2p}, this dramatic simplification does not carry over to the external entropy flow Eq.~(\ref{eq:external_en_prod})
\begin{align}
    \entropyFlow(t) &= - \frac{N}{2} \sum_{jk} \sum_\nvec \bar{P}_{\bar{\nvec}_j}(t) (P_j(t) w_{jk} - P_k(t) w_{kj}) \ln \left( \frac{n_j w_{jk}}{(n_k+1)w_{kj}} \right) \nonumber \\
    &= - \frac{N}{2} \sum_{jk} (P_j(t) w_{jk} - P_k(t) w_{kj}) \ln \left( \frac{ w_{jk}}{w_{kj}} \right) \nonumber \\
    &\phantom{m}- \frac{N}{2} \sum_{jk} \sum_\nvec \bar{P}_{\bar{\nvec}_j}(t) (P_j(t) w_{jk} - P_k(t) w_{kj}) \ln \left( \frac{n_j}{n_k +1} \right)\,, \label{eq:Se_Nid}
\end{align}
where of the last two terms only the first is the $N$-fold entropy flow of the single particle system $\entropyFlow(t)$, Eq.~(\ref{eq:se_single_part}). The reason for the second term is the lack of a cancellation mechanism to absorb the $n_j$ and $n_k + 1$ from the logarithm. Rewriting the second term as
\begin{align}\elabel{generalised_extra_term_Nid}
    &-\frac{N}{2} \sum_{jk} \sum_\nvec \bar{P}_{\bar{\nvec}_j}(t) (P_j(t) w_{jk} - P_k(t) w_{kj}) \ln \left( \frac{n_j}{n_k +1} \right) \nonumber \\
    =& - \frac{1}{2} \sum_{\nvec} \sum_{jk} \left( P_\nvec(t) n_j w_{jk} - P_\nvec \left( \frac{P_k(t) n_j}{P_j(t) (n_k+1)} \right) (n_k +1 )w_{kj}  \right) \ln \left( \frac{n_j}{n_k + 1} \right) \\
    =& - \sum_\textbf{n} \dot{P}_\textbf{n}(t) \ln \left[ {N \choose n_1,...,n_d} \right] ~,
\end{align}
using \Eref{ME_Nindindd} where we re-expressed the logarithm as 
\begin{equation}
    \ln \left( \frac{n_j}{n_k + 1} \right) = \ln \left[ 
    {N \choose n_1,\ldots,n_j-1,\ldots,n_k+1,\ldots,n_d} \right] - \ln
    \left[ 
    {N \choose n_1,\ldots,n_d}
    \right] ~,
\end{equation} 
shows that the correction term has the same form as the corresponding term in the two-state system, Eq.~(\ref{eq:boltz_contrib_se}), namely that of a difference of microcanonical (Boltzmann) entropies of the multi-particle states.
It vanishes when all $n_j$ are either 0 or 1, as expected for $d \gg N$ and also at stationarity when $\dot{P}_\textbf{n}(t) = 0$. In that limit $\entropyFlow=-\entropyProduction$ when indeed \Eref{si_single_part} gives
\begin{equation}
\lim_{t\to\infty}
    \entropyProduction^{(1)}(t) = 
    \frac{1}{2} \sum_k (P_j w_{jk} - P_k w_{kj}) \ln \left( \frac{w_{jk}}{w_{kj}} \right) ~,
\end{equation}
with $P_j = \lim_{t\to\infty} P_j(t)$.
As far as the entropy production $\entropyProduction(t)$ is concerned, we thus recover and generalise the result in Sec.~\ref{sec:N2p} on indistinguishable particles in a two-state system, which produce $N$ times the entropy of a single particle. In Sec.~\ref{sec:Ndist} it was shown that $N$ \textit{distinguishable} particles have the same entropy production and flow as the sum of the entropy productions of individual particles. In Sec.~\ref{sec:N2p} and \ref{sec:Nindepd} it was shown that the entropy production of \textit{indistinguishable} particles, which require the states to be represented by occupation numbers, show the $N$-fold entropy production of the single particle system, provided suitable initialisation, but asymptotically independent of initialisation, provided the stationary state has a unique distribution. The same does not apply to the entropy flow, which generally acquires additional logarithmic terms accounting for the degeneracy of the occupation number states. The extra terms, however, are bound to vanish at stationarity, when $\entropyFlow(t) = - \entropyProduction(t)$.


\subsection{Random Walk on a lattice}
\label{sec:CTRW1d}

\begin{figure}[h]
\centering
\includegraphics[width=0.8\textwidth]{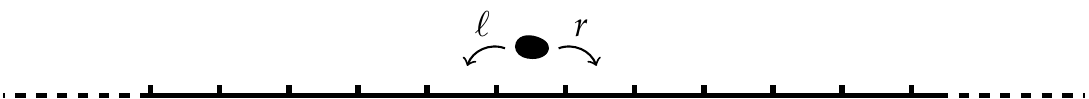}
\caption{\flabel{F7} Simple random walk on an infinite, one-dimensional lattice in continuous time. The black blob indicates the current position of the random walker. The left and right hopping rates, labelled $\ell$ and $r$ respectively, are assumed to be homogeneous but not equal in general, this leading to a net drift of the average position.}
\end{figure}   

In this section we study 
a particle on a one-dimensional lattice that hops to the  
right nearest neighbouring site with rate $r$ and to the left with rate $\ell$, see \fref{F7}. The position $x$
of the particle at time $t$, after $N(t)$ jumps, is
\begin{equation}\elabel{x_after_N_jumps}
    x=x_0+\sum_{i=1}^{N(t)}\Delta x_i,
\end{equation}
where the random hops $\Delta x_i$ are independent and identically distributed,
and $x_0$ is the initial position at time $t=0$.
If $\latticespacing$ is the lattice spacing, the distance increments are 
$\Delta x_i=+\latticespacing$ with probability $r/(\ell+r)$ and
$\Delta x_i=-\latticespacing$ with probability $\ell/(\ell+r)$.
The probability distribution of the particle position is
\begin{equation}
    \elabel{PXt_def}
    P(x,t;x_0)=\sum_{n=0}^{\infty}H(n,t)P_n(x;x_0)\,,
\end{equation}
where $H(n,t)$ is the probability that by time $t$, the particle has hopped
$N(t)=n$ times, and
$P_n(x;x_0)$ is the probability that the particle is at position $x$
after $n$ hops starting from $x_0$. Since  jumping is a Poisson process
with rate $r+\ell$, the random variable $N(t)$ has a Poisson distribution,
\begin{equation}
    H(n,t)=\frac{((\ell+r)t)^n}{n!}\exp{-(\ell+r)t} \,.
    \elabel{Pnt}
\end{equation}
On the other hand, the distribution of the position $x$ after $n$ 
jumps is the binomial distribution
\begin{equation}
    P_n(x;x_0) 
    = \binom{n}{k_x}
    \frac{r^{k_x}\ell^{n-k_x}}{(\ell+r)^n}\,,
    \elabel{Pnx}
\end{equation}
where $k_x=(n+(x-x_0)/\latticespacing)/2$ is the number of jumps to the right,
$0\leq k_x\leq n$ with \eref{Pnx} implied to vanish if $k_x$ is not
integer. 
From \Eref{x_after_N_jumps} the parity of $(x-x_0)/\latticespacing$ and $N(t)$ are identical.
Using \eref{Pnt} and \eref{Pnx},
the probability distribution in \eref{PXt_def} reads
\begin{align}
    P(x,t;x_0) 
    =& \exp{-(\ell+r)t} \left(\frac{r}{\ell}\right)^{\frac{x-x_0}{2\latticespacing}}
    \IC\left(\frac{|x-x_0|}{\latticespacing},2t\sqrt{r\ell}\right)
    \,,
    \elabel{PXt_res}
\end{align}
where $\IC(n,z)$ is the modified Bessel function of the first kind.\footnote{The
modified Bessel function of the first kind of $m,z\in\mathbb{C}$ is defined as
\cite{MagnusOberhettingerSoni:1966}
\begin{equation}
\IC(m,z) = \sum_{j=0}^\infty \frac{1}{j!\Gamma(j+m+1)}\left(\frac{z}{2}\right)^{2j+m} \,.
\elabel{defBesselI}
\end{equation}
}
The transition probability is then,
\begin{equation}
W(x\to y;\tau) = \exp{-(\ell+r)\tau} \left(\frac{r}{\ell}\right)^{\frac{y-x}{2\latticespacing}}
    \IC\left(\frac{|y-x|}{\latticespacing},2\tau\sqrt{r\ell}\right) \,.
    \elabel{WCTRW}
\end{equation}
Using \eref{PXt_res} and \eref{WCTRW} to calculate the entropy production \eref{internal_en_prod}, we need the following identity for $|y-x|/\latticespacing=|m|\geq1$,
\begin{equation}
\lim_{\tau\to0}\frac{ 1}{\tau}  
\IC\left(|m|,2 \tau\sqrt{r\ell}\right) 
= \sqrt{r\ell} \delta_{|m|,1} 
\,,
\elabel{lim_BesselI}
\end{equation}
which follows immediately from \Eref{defBesselI}. It indicates that the only transitions that contribute to the entropy production are those where the particle
travels a distance equal to the lattice spacing $\latticespacing$. Then, the entropy production reads,
\begin{align}
\entropyProduction(t) = &
\frac{1}{2}(r-\ell)\ln\left(\frac{r}{\ell}\right) 
+
\exp{-(\ell+r) t} \sum_{m=-\infty}^\infty  \left(\frac{r}{\ell}\right)^{\frac{m}{2}} 
    \IC\left(|m|,2t\sqrt{r\ell}\right) \nonumber\\& \times
\left[
 {r}\ln{ \left(\frac{\IC\left(|m|,2t\sqrt{r\ell}\right)}{\IC\left(|m+1|,2t\sqrt{r\ell}\right)}\right) }
 +{\ell}\ln{ \left(\frac{\IC\left(|m|,2t\sqrt{r\ell}\right)}{\IC\left(|m-1|,2t\sqrt{r\ell}\right)}\right) }
\right] \,.
\elabel{ent_prod_CTRW_expr}
\end{align}
and the entropy flow $\entropyFlow(t) =-(r-\ell)\ln(r/\ell)$ independent of $t$, which owes its simplicity to the transition rates being independent of the particle's position. 
{We are not aware of a method 
to perform the sum in
\eref{ent_prod_CTRW_expr} in closed form and, given that
this expression involves terms competing at
large 
times $t$, we cannot calculate the stationary entropy
production $\lim_{t\to\infty}\entropyProduction(t)$. 
If we assume that the
sum in \Eref{ent_prod_CTRW_expr} converges such that the exponential $\exp{-(r+\ell)t}$ 
eventually suppresses it, then the entropy production $\entropyProduction$ appears to converge to
$\frac{1}{2}(r-\ell)\ln(r/\ell)$. 
 If that were the 
 case, $\dot{S}=\entropyProduction+\entropyFlow$
would converge to a negative constant, while $S(t)$,
\Eref{entropy}, which vanishes at $t=0$ given the initialisation of $P(x,t;x_0)=\delta_{({x-x_0})/{\latticespacing},0}$, is 
bound to be strictly positive at all finite $t$. Given
that $P(x,t;x_0)$ does not converge, not much else can
be said about $S(t)$ or $\dot{S}$.}
Using the master
equation
\begin{equation}
    \dot{P}(x,t;x_0) = 
    -(r+\ell) P(x,t;x_0)
    +\ell P(x+\latticespacing,t;x_0)
    -r P(x-\latticespacing,t;x_0)
\end{equation}
in
\begin{subequations}
\begin{align}
    \dot{S}(t) = &-\sum_m 
    \dot{P}(m\latticespacing,t;x_0)\ln(P(m\latticespacing,t;x_0))\\
    =&\sum_m \{(r+\ell)P(m\latticespacing,t;x_0) 
    -\ell P((m+1)\latticespacing,t;x_0)
    -r P((m-1)\latticespacing,t;x_0)\} \ln(P(m\latticespacing,t;x_0)) \elabel{RW_half_attempt_b}\\
    =&\sum_m rP(m\latticespacing,t;x_0)\ln\left(\frac{P(m\latticespacing,t;x_0)}{P((m+1)\latticespacing,t;x_0)}\right)
    +\ell P(m\latticespacing,t;x_0) \ln\left(\frac{P(m\latticespacing,t;x_0)}{P((m-1)\latticespacing),t;x_0}\right) \elabel{RW_half_attempt_c}
\end{align}    
\end{subequations}
still requires an approximation such as
the continuum limit in
\Eref{Gaussian_contlimitRW} either in the logarithm of the
ratios in \Eref{RW_half_attempt_c} or in the logarithm of 
$P(m\latticespacing,t;x_0)$ in \Eref{RW_half_attempt_b}.
The resulting sum can be performed elegantly using, for
example, $\sum_m P(m\latticespacing,t;x_0)m - P((m+1)\latticespacing,t;x_0)(m+1)=0$. Remarkably, either
approach produces $-\frac{1}{2}(r-\ell)\ln(r/\ell)$ for
$\dot{S}$.
Using $\latticespacing/\sqrt{t}$ as the 
integration mesh, the sum can be 
re-interpreted as a Riemann sum and
the difference in the summand 
Taylor expanded to give $\dot{S} =0$ in
large $t$. Even when this result
is more reasonable than negative 
$\lim_{t\to\infty}\dot{S}(t)$, we are not aware of a 
rigorous proof that $\lim_{t\to\infty}\dot{S}(t)=0$, and thus not of a proof
of the corresponding limit 
$\lim_{t\to\infty}\dot{S}(t)=(r-\ell)\ln(r/\ell)$. The
closely related Brownian particle,
discussed in \sref{DDparticle}
does not suffer from this 
difficulty.

To take the continuum limit $\latticespacing\to0$ of the probability
distribution \eref{PXt_res},
we define $\drift$ and $D$ such that $r+\ell= 2D/\latticespacing^2$
and $r-\ell=\drift/\latticespacing$. Using the asymptotic expansion\footnote{
We use the asymptotic expansion in $m$ of the modified Bessel 
function \cite{MagnusOberhettingerSoni:1966}
\begin{equation}
    \IC(m,z) \sim \frac{\exp{z}}{\sqrt{2\pi z}}
    \left( 1- \frac{4m^2-1}{8z}
    +\frac{\left(4m^2-1\right)\left(4m^2-9\right)}{2!(8z)^2}
    -\frac{\left(4m^2-1\right)\left(4m^2-9\right)\left(4m^2-25\right)}{3!(8z)^3}
    +\ldots
    \right)\,,
\end{equation}
which is valid for $|\arg z|<\pi/2$.
} of $\IC(m,z)$ in $m$, we obtain in fact the Gaussian distribution,
\begin{equation}
    \lim_{\latticespacing\to0} \frac{1}{\latticespacing}
    P\left(\frac{x}{\latticespacing},t;\frac{x_0}{\latticespacing};r(\drift,D,\latticespacing),\ell(\drift,D,\latticespacing)\right)
    = \frac{1}{\sqrt{4\pi Dt}}\exp{-\frac{(x-x_0-\drift t)^2}{4Dt}} \,,
    \elabel{Gaussian_contlimitRW}
\end{equation}
which corresponds to
the distribution of a drift-diffusive particle, which is
studied in \sref{DDparticle}.
Therefore, all results derived in \sref{DDparticle}, apply 
to the present system in the continuum limit.

\subsection{Random Walk on a ring lattice}
\label{sec:RWring}

\begin{figure}[h]
\centering
\includegraphics[width=0.25\textwidth]{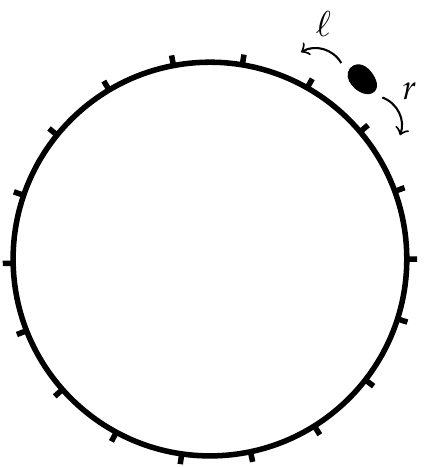}
\caption{\flabel{F8} Simple random walk on an periodic, one-dimensional `ring' lattice in continuous time. This model generalises the three-state Markov chain discussed in Section \ref{sec:sec3state} to $L$ states. The black blob indicates the current position of the random walker. Due to the finiteness of the state space, this process is characterised by a well defined steady-state, which is an equilibrium one for symmetric rates $\ell = r$. }
\end{figure}   

In this section we extend the system in \sref{CTRW1d} to a random walk on a 
ring lattice of length $L>2$, so that $1\leq x\leq L$, see \fref{F8}.
The probability distribution $P_L(x,t)$ of the particle on the ring follows
from the distribution on the one-dimensional lattice $P(x,t)$ in 
\eref{PXt_res},
by mapping all positions $x+jL$ on the one-dimensional
lattice to position $x\in\{1,2,\ldots,L\}$ on the ring
with $j$ being the winding number irrelevant to the 
evolution of the walker.
Then, the distribution on the ring lattice reads,
\begin{equation}
    P_L(x,t;x_0) = \sum_{j=-\infty}^{\infty}
    P(x+jL,t;x_0) \,
    \elabel{distr_RWring}
\end{equation}
and similarly for the transition probability $W(x\to y,\tau)=P_L(y,\tau;x)$.
To calculate the entropy production \eref{internal_en_prod}, each pair of points $x,y$
on the lattice is mapped to a pair of points on the ring.
For $L>2$, as $\tau\to0$ only 
transitions to distinct, nearest neighbours 
contribute and the expression
for the entropy production 
simplifies dramatically,
\begin{equation}
    \entropyProduction(t) = (r-\ell)\ln\left(\frac{r}{\ell}\right)
    +\sum_{m=1}^{L/\latticespacing} P_L(m\latticespacing,t;x_0)\left\{
    r \ln\left(\frac{P_L(m\latticespacing,t;x_0)}{P_L((m+1)\latticespacing,t;x_0)}\right)
    + \ell \ln\left(\frac{P_L(m\latticespacing,t;x_0)}{P_L((m-1)\latticespacing,t;x_0)}\right)
    \right\}
    \elabel{Si_ring_lat}
\end{equation}
and similar for 
\begin{equation}
    \entropyFlow(t) = -(r-\ell)\ln\left(\frac{r}{\ell}\right).
\end{equation}
While the entropy flow $\entropyFlow$ on a ring is thus identical to that of a particle on a 
one-dimensional lattice, the entropy production $\entropyProduction$ on a ring is in principle
more complicated, but with a lack of cancellations of $\sqrt{r/\ell}$ in the logarithm as found
in Sec.~\ref{sec:CTRW1d} and $P_L$ reaching stationarity comes the asymptote
\begin{equation}
\label{eq:steady_Si_ring_lattice}
    \lim_{t\to\infty}\entropyProduction(t) = (r-\ell) \ln\left(\frac{r}{\ell}\right).
\end{equation}
This is easily derived from $\lim_{t\to\infty}P_L(x,t;x_0)=1/L$ taken into the
finite sum of \Eref{Si_ring_lat}. It follows that $\dot{S}(t)=\entropyProduction(t)+\entropyFlow(t)$ converges to $0$ at large $t$,
as expected for a convergent stationary distribution.

The case $L=2$ and the less interesting case $L=1$ are not covered above, because of the different
topology of the phase space of $L>2$ compared to $L=2$.
The difference can be observed in the different structure of the transition matrices 
\eref{W2} and \eref{transmat3}. The framework above is based on each site having two outgoing
and two incoming rates, $2L$ in total. However, for $L=2$ there are only two transitions,
which cannot be separated into four to fit the framework above, because even when rates of 
concurrent transitions between two given states are additive, their entropy production generally
is not. The case of $L=2$ is recovered in the two-state system of \sref{sec2state} with 
$\alpha=\beta=r+\ell$, which is at equilibrium in the stationary state.



\subsection{Driven Brownian particle}
\label{sec:DDparticle}
\begin{figure}[h]
\centering
\includegraphics[width=0.7\textwidth]{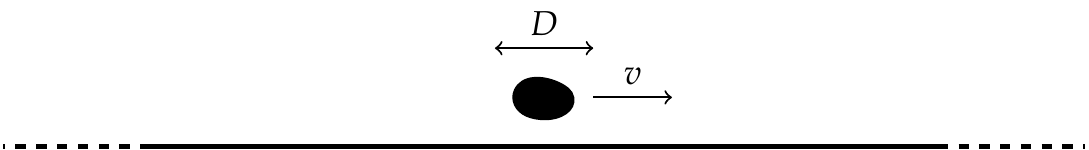}
\caption{\flabel{F9} Driven Brownian particle on the real line. The black blob indicates the particle's current position.}
\end{figure} In continuum space,
the motion of a freely diffusive particle with diffusion constant $D$ and drift $\drift$ is governed by the 
 Langevin equation
$\dot{x} = \drift+\sqrt{2D}\xi(t)$,
where  $\xi(t)$ is a Gaussian white noise with zero mean, $\ave{\xi(t)}=0$, and covariance $\ave{\xi(t)\xi(t')}=\delta(t-t')$, see \fref{F9}
\cite{Van:2010}.
The corresponding Fokker-Planck equation for the probability distribution $P(x,t)$ is \cite{Risken1996}
\begin{equation}
\label{eq:FP_driven_BP_line}
\partial_t P(x,t)= -\drift \partial_x P(x,t) + D\partial^2_x P(x,t) \, .
\end{equation}
Assuming the initial condition $P(x,0)=\delta(x-x_0)$, the solution to the Fokker-Planck equation is the Gaussian distribution
\begin{equation}
P(x,t) = \frac{1}{\sqrt{4D\pi t}}\exp{-\frac{(x-x_0 -\drift t)^2}{4D t}} \, ,
\elabel{DDPx}
\end{equation}
which is also the Green function of the Fokker-Planck equation (\ref{eq:FP_driven_BP_line}). We therefore also have the transition probability density from state x to state y over an interval $\tau$,
\begin{equation}
W(x\to y,\tau) = \frac{1}{\sqrt{4D\pi\tau}}\exp{-\frac{(y-x -\drift\tau)^2}{4D\tau}} \, .
\elabel{DDWx}
\end{equation}
Substituting \eref{DDPx} and \eref{DDWx} into  \Eref{continuum_en_prod} for the internal entropy production of a continuous system
gives,
\begin{equation}
\elabel{int_en_prod_dd}
\entropyProduction = \lim_{\tau\to0} \frac{1}{\tau}
\int\dint{x}\dint{y}
\frac{1}{\sqrt{4D\pi t}}
\exp{-\frac{(x -x_0 -\drift t)^2}{4Dt}}
\frac{1}{\sqrt{4D\pi\tau}}
\exp{-\frac{(y-x -\drift \tau)^2}{4D\tau}}
\left(\frac{(y-x_0)^2-(x-x_0)^2}{4Dt} + \frac{(y-x)\drift }{2D}\right) \,,
\end{equation}
where the Gaussian integrals can be evaluated in closed form,
$\entropyProduction(t) =  \lim_{\tau\to0} \left[ {1}/({2t}) +{\drift^2}/{D}+{\drift^2}\tau/({2Dt}) \right]
$. Taking the limit
$\tau\to0$ then gives the entropy production rate \cite{Van:2010,Maes:2000,Spinney:2012},
\begin{equation}
\entropyProduction(t) =   \frac{1}{2t} +\frac{\drift^2}{D}\,.
\elabel{siDD}
\end{equation}
Similarly, following \eref{continuum_en_flow},
 the entropy flow  reads $\entropyFlow(t)=-\drift^2/D$ 
 independent of time $t$. 
As $\entropyProduction(t)\ne0$,
we see that for finite $t$ or $\drift\neq0$, the system is out of equilibrium with a sustained probability current, so that there is 
in fact no steady-state distribution.
We can verify 
\Eref{siDD} for the time-dependent internal entropy production by computing the probability current 
\begin{equation}
    j(x,t) = (\drift - D \partial_x) P(x,t) =  \left( \frac{\drift }{2 }+\frac{(x-{x_0}-\drift t)}{4 t } \right) \frac{e^{-\frac{(x-x_0- \drift t)^2}{4 D t}}}{ \sqrt{\pi  D t}}
\end{equation}
and substituting it together with \eref{DDPx}, into (\ref{eq:tot_en_prod}). As expected, the two procedures return identical results.
The independence of the transient contribution $1/(2t)$ to the internal entropy production on the diffusion constant is remarkable although necessary on  dimensional grounds, as a consequence of $\entropyProduction$ having dimensions of inverse time. The diffusion constant characterising the spatial behaviour of diffusion suggests that it is the temporal, rather than spatial features of the process that determine its initial entropy production.

\subsection{Driven Brownian particle in a harmonic potential}
\label{sec:OU}
\begin{figure}[h]
\centering
\includegraphics[width=0.4\textwidth]{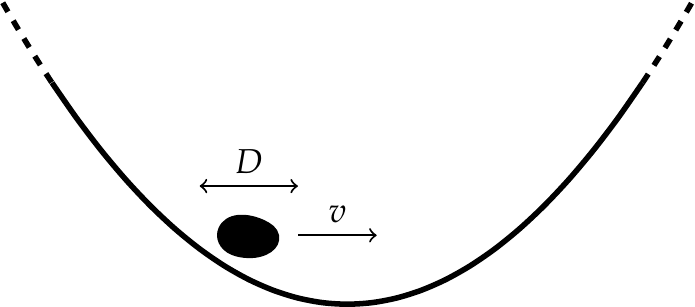}
\caption{\flabel{F10} Driven Brownian particle in a harmonic potential. This process reduces to the standard Ornstein-Uhlenbeck process upon rescaling $x \to x' + \drift/k$.  The black blob indicates the particle's current  position. The presence of a binding potential implies that the system relaxes to an equilibrium steady-state at long times.}
\end{figure} 
Consider a drift-diffusive particle such as in \Sref{DDparticle} that is confined in a harmonic potential 
$V(x)=\frac{1}{2}kx^2$, where $k$ is the potential stiffness, see \fref{F10}
\cite{Andrieuxetal:2008}. The Langevin equation is  
$\dot{x} = \drift -kx +\sqrt{2D}\xi(t)$,
where $\ave{\xi(t)}=0$ and $\ave{\xi(t)\xi(t')}=\delta(t-t')$ and the Fokker-Planck equation for $P(x,t)$ is 
\cite{Risken1996}
\begin{equation}
\partial_t P(x,t)= - \partial_x ((\drift-kx)P(x,t)) + D\partial^2_x P(x,t) \, .
\end{equation}
Assuming the initial condition $P(x,0)=\delta(x-x_0)$, the solution
to the Fokker-Plank equation is the Gaussian distribution 
\begin{equation}
P(x,t) = \sqrt{\frac{k}{2\pi D(1-\exp{-2kt})}}\exp{-\frac{\left(kx-\drift-\left(kx_0-\drift\right)\exp{-kt}\right)^2}{2 Dk(1-\exp{-2kt})}} \, ,
\elabel{OUdP}
\end{equation}
corresponding to a probability current $j(x,t)=(\drift-kx-D\partial_x)P(x,t)$ of the form
\begin{equation}
    j(x) = \frac{\sqrt{\frac{k}{2 \pi  D ( 1- e^{-2 k t})}} e^{-k t}\left(\drift \left(1 - e^{-k t}\right)- k \left(x_0-x e^{-k t}\right)\right) }{1-e^{-2 k t}} 
    \exp{ -\frac{(\drift (1- e^{-k t} ) - k (x - x_0 e^{-k t})^2}{2 D k (1 - e^{-2 k t})}} ~.
\end{equation}
The transition probability density within $\tau$ is then also of Gaussian form, namely
\begin{equation}
W(x\to y,\tau) = \sqrt{\frac{k}{2\pi D(1-\exp{-2k\tau})}}\exp{-\frac{\left(ky-{\drift}-\left(kx-{\drift}\right)\exp{-k\tau}\right)^2}{2 Dk(1-\exp{-2k\tau})}} \, .
\elabel{OUdW}
\end{equation}
Using \eref{OUdP} and \eref{OUdW} in \eref{continuum_en_prod} gives the entropy production rate
\begin{equation}
\entropyProduction =  \left(\frac{(\drift-kx_0)^2}{D}-k\right)\exp{-2kt}+\frac{k\exp{-2kt}}{1-\exp{-2kt}}\,
\elabel{OUsiG}
\end{equation}
and in (\ref{eq:continuum_en_flow}) the external entropy flow
\begin{equation}
    \entropyFlow=-\left(\frac{(\drift-kx_0)^2}{D}-k\right)\exp{-2kt}\,.
    \elabel{Se_harm}
\end{equation}
In the limit $t\to\infty$, the system will reach equilibrium as $P(x,t)$ in Eq.~\eref{OUdP} converges to the Boltzmann distribution $\sqrt{\frac{k}{2\pi D}}\exp{-\frac{(kx-\drift)^2}{2Dk}}$ of the effective potential $\frac{1}{2}kx^2-\drift x$ at temperature $D$. This is consistent with \eref{OUsiG} and \eref{Se_harm} since $\lim_{t\to\infty}\entropyProduction(t)=\lim_{t\to\infty}\entropyFlow(t)=0$. Similarly to drift diffusion on the real line, \Eref{siDD}, there is a transient contribution to the entropy production that is independent of the diffusion constant $D$ but does now depend on the stiffness $k$, which has dimensions of inverse time, through the rescaled time $kt$.

\subsection{Driven Brownian particle on a ring with potential}
\label{sec:Bpartring}
\begin{figure}[h]
\centering
\includegraphics[width=0.4\textwidth]{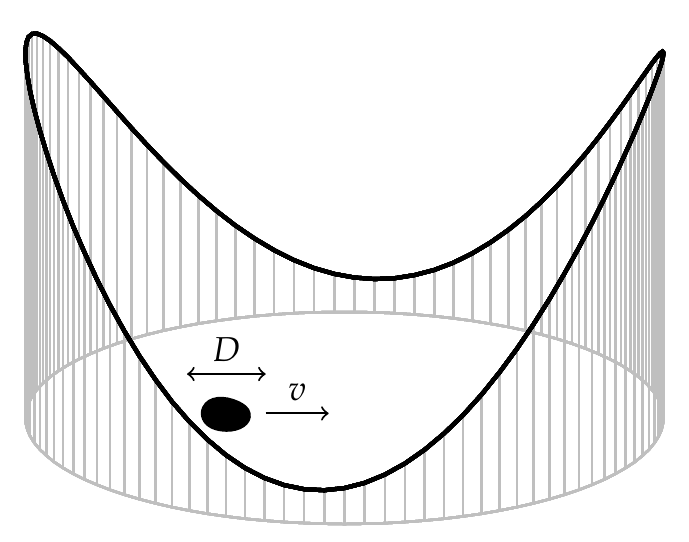}
\caption{\flabel{F11} Driven Brownian particle on a ring $x\in[0,L)$ with a periodic potential satisfying $V(x)=V(x+L)$. Any finite diffusion constant $D>0$ results in a stationary state at long times that is non-equilibrium for $v\ne0$. The black blob indicates the particle's current position. }
\end{figure} 
Consider a drift-diffusive particle on a ring $x\in[0,L)$ in a smooth potential $V(x)$,
 \fref{F11}, initialised at position $x_0$.
The Langevin equation  of the particle is \cite{Reimann2001Jun,Pigolotti:2017,Neri:2019}
$\dot x = \drift -\partial_xV(x)+\sqrt{2D}\xi(t)$,
where $\xi(t)$ is Gaussian white noise.
The Fokker-Planck equation is then 
\begin{equation}
\partial_t P(x,t;x_0) = - \partial_x ((\drift - V'(x))P(x,t;x_0)) +D\partial^2_x P(x,t;x_0)
\elabel{FPpotring}
\end{equation}
 with $V'(x)=\frac{d}{dx}V(x)$ and boundary condition $P^{(n)}(0,t;x_0)=P^{(n)}(L,t;x_0)$ for for all $n\ge0$ derivatives and $t\geq0$.
 At stationarity, in the limit $t\to\infty$, where $\partial_t P(x,t;x_0) = 0$,
the solution to the Fokker-Planck equation \eref{FPpotring} is \cite{Horsthemke1984, Pavliotis2014,Neri:2019}
\begin{equation}
P_s(x)=\lim_{t\to\infty}P(x,t)={\ZC} \exp{-\frac{V(x)-\drift x}{D}} \int^{x+L}_{x}\dint{y} \exp{\frac{V(y)-\drift y}{D}} \,,
\elabel{statsol}
\end{equation}
where $\ZC$ is the normalisation constant. The corresponding steady-state probability current $j=(\drift-\partial_xV)P_s-D\partial_xP_s$ is independent of $x$ by continuity, $0=\partial_t P=-\partial_x j$, and reads \cite{Risken1996}
\begin{align}
    j &= \ZC\left( \exp{-\frac{\drift L}{D}}-1\right)
    \label{current_drift_pot} ~.
\end{align}

In order to calculate the entropy production according to 
(\ref{eq:continuum_en_prod}) and 
(\ref{eq:continuum_en_flow})
using (\ref{eq:hopital_continuum}), we need $W(x\to y;\tau)$ for small $\tau$. As discussed after
\Eref{hopital_continuum}, $W(x\to y;\tau)$ obeys the Fokker-Planck
\Eref{FPpotring} in the form 
\begin{equation}
\partial_{\tau}W(x\to y;\tau)=-\partial_y\left[ \left(\drift-V'(y)\right)W(x\to y;\tau)\right]+D\partial_y^2W(x\to y;\tau)
\end{equation}
with $\lim_{\tau\to 0} W(x\to y;\tau)=\delta(y-x)$, 
so that
\begin{equation}
    \dot{W}(x\to y)=\lim_{\tau \to 0}\partial_\tau W(x\to y;\tau)=V''(y)\delta(y-x)-(\drift-V'(y))\delta'(y-x)+D\delta''(y-x) \label{eq:Wdot_op}
\end{equation}
 to be evaluated under an integral, where $\delta'(y-x)=\frac{d}{dy}\delta(y-x)$ will require an integration by parts. As for the logarithmic term, we use 
 \cite{Wissel1979Jun, Risken1996}
\begin{equation}
    W(x\to y;\tau)=\frac{1}{\sqrt{4\pi D\tau}}e^{-\frac{\left(y-x-\tau\left(\drift-V'(x)\right)\right)^2}{4D\tau}}\left(1+\mathcal O(\tau)\right)
\end{equation}
so that 
\begin{equation}
    \ln\left( \frac{W(x\to y;\tau)}{W(y\to x;\tau)} \right)=\frac{y-x}{2D}\left(2\drift-V'(x)-V'(y)\right)+\mathcal O(\tau).
    \label{eq:Log_ring_BP}
\end{equation}
The entropy flow \Eref{continuum_en_flow}
in the more convenient version
\Eref{Se_cont_easy}
can be obtained easily  using 
\Erefs{Wdot_op} and \eref{Log_ring_BP},
\begin{align}
    \entropyFlow(t)&=-\int_0^L \dint{x} \dint{y} P(x,t)\Big(V''(y)\delta(y-x)-(\drift-V'(y))\delta'(y-x)+D\delta''(y-x)\Big)\\
    & \qquad\qquad \times\frac{y-x}{2D}\left( 2\drift-V'(x)-V'(y)\right)\\
    &=-\int_0^L \dint{x} P(x,t)\left(\frac{1}{D}\left(\drift-V'(x)\right)^2-V''(x) \right)
    \label{eq:Se_ring_BP}
\end{align}
after suitable integration by parts, whereby derivatives of the $\delta$-function are conveniently interpreted as derivatives with respect to $y$ to avoid subsequent differentiation of $P(x,t)$. Since $\delta(y-x)(y-x)=0$, the factor $(y-x)/(2D)$ needs to be differentiated for a term to contribute. In the absence of a potential, $P(x,t)=1/L$ at stationarity, so that \Eref{Se_ring_BP} simplifies to $\entropyFlow(t)=-{\drift^2}/{D}$ and 
$\lim_{t\to\infty}\entropyProduction(t)=\drift^2/D$, \Eref{siDD}.
Using the probability current $j(x,t)=-D\partial_xP(x,t)+\left( \drift-V'(x)\right)P(x,t)$, the entropy flow simplifies further to 
\begin{equation}
    \entropyFlow(t)=-\int_0^L \dint{x} j(x,t)\frac{\drift-V'(x)}{D}
\end{equation}
so that at stationarity, when the current is spatially uniform, $\lim_{t \to\infty}\entropyFlow(t)=-
\lim_{t\to\infty}j(x,t)\drift L/D$ as the potential is periodic,
entering only via the current.

An equivalent calculation of $\entropyProduction$ on the basis of (\ref{eq:Si_cont_easy}) gives
\begin{subequations}
\begin{align}
    \entropyProduction(t)&=-\entropyFlow+\int_0^L \dint{x} \dint{y} P(x,t)\Big(V''(y)\delta(y-x)-(\drift-V'(y))\delta'(y-x) +D\delta''(y-x)\Big)\ln\left(\frac{P(x,t)}{P(y,t)} \right)\\
    &=-\entropyFlow+\int_0^L \dint{x} \left\{D \frac{(P'(x,t))^2}{P(x,t)}-P(x,t)V''(x)\right\}\\
    &=-\entropyFlow-\int_0^L \dint{x} j(x,t)\partial_x \ln P(x,t)\\
    &=\int_0^L \plaind x \frac{j^2(x,t)}{DP(x,t)}\,, \label{eq:Si_ring_BP_j}
\end{align}
\end{subequations}
with the last line identical to
\Eref{continuous_entropies_decomp_prod}.

By considering the functional derivative $\delta Z/\delta V(z)$ in $\int^L_0 \plaind x P(x)=1$ of \Eref{statsol},
one can show that the stationary current $j(x,t)$ Eq.~(\ref{current_drift_pot}) is extremal for constant $V(x)$, indicating that the magnitude of the stationary entropy flow \Eref{Si_ring_BP_j} is maximised in a constant potential.

\subsection{Run-and-tumble motion with diffusion on a ring}
\label{sec:RnTring}
\begin{figure}[h]
\centering
\includegraphics[width=0.4\textwidth]{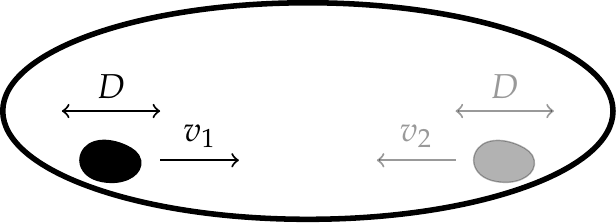}
\caption{\flabel{F12} Run-and-tumble motion with diffusion on a ring $x \in [0,L)$. A run-and-tumble particle switches stochastically, in a Poisson process with rate $\alpha$, between two modes $1$ and $2$ characterised by an identical diffusion constant $D$ but distinct drift velocities $\drift_1$ and $\drift_2$. The two modes are here represented in black and grey, respectively. For arbitrary positive diffusion constant $D$ or tumbling rate $\alpha$ with $\drift_1\ne\drift_2$ the steady state is uniform but generally non-equilibrium.}
\end{figure} 
Consider the dynamics of a  run-and-tumble particle on a ring $x\in [0,L)$ \cite{Schnitzer1993Oct} with Langevin equation
$\dot x=\drift(t)+\sqrt{2D}\xi(t)$,
where the drift $\drift(t)$ is a Poisson process with rate $\alpha$ that alternates 
the speed of the particle between the constants
 $\drift_1$ and $\drift_2$, and $\xi(t)$ is Gaussian white noise, \fref{F12}. The drift being $\drift(t)=\drift_1$ or $\drift(t)=\drift_2$ will be referred to as the mode of the particle being $1$ or $2$ respectively. Defining $P_1(x,t)$ and $P_2(x,t)$
  as the joint  probabilities that the particle is at position $x$ at time $t$ and in mode $1$ or $2$
  respectively, the coupled Fokker-Planck equations for $P_1$ and $P_2$ are
\begin{subequations}
\label{eq:system_FP_rnt}
\begin{align}
\partial_tP_1(x,t)= & -\drift_1\partial_x P_1(x,t)+D\partial_x^2 P_1(x,t) -\alpha(P_1(x,t)-P_2(x,t))\elabel{rntfp1}\\\elabel{rntfp2}
\partial_tP_2(x,t)= & -\drift_2\partial_x P_2(x,t)+D\partial_x^2 P_2(x,t)-\alpha(P_2(x,t)-P_1(x,t))\,
\end{align}
\end{subequations}
whose stationary solution is the uniform distribution $\lim_{t\to\infty}P_1(x,t)=\lim_{t\to\infty}P_2(x)=1/(2L)$ as is easily verified by direct substitution. The corresponding steady-state probability currents thus read $j_1 = \drift_1/(2L) $ and $j_2 = \drift_2/(2L) $.

In the following, we denote by the propagator $W(x\to y,Q\to R;\tau)$ the probability density that a particle at position $x$ in mode $Q$ is found time $\tau$ later at position $y$ in mode $R$. For $Q=R$, this propagation is a sum over all  even numbers $m$ of Poissonian switches, that occur with probability $(\alpha\tau)^m\exp{-\alpha \tau}/m!$, which includes the probability $\exp{-\alpha \tau}$ of not switching at all over a total of time $\tau$. For $Q\neq R$, the propagation is due to an odd number of switches.

For $m=0$, the contribution to $W(x\to y,Q\to R;\tau)$ is thus $\exp{-\alpha \tau}W(x\to y;\tau)$, with $W(x\to y;\tau)$ of a drift diffusion particle on a ring, \Sref{Bpartring}, but without potential, approximated at short times $\tau$ by the process on the real line, \Eref{DDWx}
 with drift $\drift=\drift_1$ or $\drift=\drift_2$ according to the particle's mode. For $m=1$ the contribution is a single convolution over the time $t'\in[0,\tau)$ at which the particle changes mode, most easily done after Fourier transforming. Before presenting this calculation in real space, we argue that any such convolution will result in some approximate Gaussian with an amplitude proportional to $1/\sqrt\tau$ multiplied by a term of order $(\alpha\tau)^m$. In small $\tau$, therefore only the lowest orders need to be kept, $m=0$ for $Q=R$ and $m=1$ for $Q\neq R$.
 
More concretely, 
\begin{align}
    &W(x\to y, 1\to 2;\tau) \nonumber\\
    =&\int^{\infty}_{-\infty}\plaind z \int^{\tau}_0 \plaind \tau'\frac{1}{\sqrt{4\pi D \tau'}}\exp{-\frac{(z-x-\drift_1\tau')^2}{4D\tau'}} \exp{-\alpha\tau'}
    \frac{1}{\sqrt{4\pi D(\tau-\tau')}}\exp{-\frac{\left(y-z-\drift_2(\tau-\tau')\right)^2}{4D(\tau-\tau')}}
    \exp{-\alpha(\tau-\tau')}+\ldots\\
    =&\frac{\alpha\exp{-\alpha\tau}}{2(\drift_1-\drift_2)}\left[\text{erf}\left(\frac{x-y+\drift_1\tau }{\sqrt{4D\tau}}\right)-\text{erf}\left(\frac{x-y+\drift_2\tau}{\sqrt{4D\tau}}\right)\right]+\ldots 
    \label{propagator}
 \end{align}
 which in small $\tau$, when $\drift_{1,2}\tau/\sqrt{4D\tau} \ll 1$, so that
 $\text{erf}(r+\varepsilon)=\text{erf}(r)+{2\varepsilon}\,e^{-r^2}/\sqrt{\pi}+\ldots$, 
 expands to 
 \begin{align}
     W(x\to y,1\to 2;\tau)&=
     \frac{\alpha\tau}{\sqrt{4\pi D \tau}}
     e^{-\frac{(y-x)^2}{4D\tau}}
     \left( 1 +\mathcal{O} (\tau^2)\right)
     \label{eq:RnT_trans_expanded}\nonumber\\
     &=W(x\to y,2\to 1;\tau)\ ,
 \end{align}
 whereas $W(x\to y,Q\to Q;\tau)$, the propagator with an even number of mode switches, is given by \Eref{DDWx} to leading order in $\tau$,
 \begin{equation}
     W(x\to y,Q\to Q;\tau) = 
      \frac{1}{\sqrt{4\pi D \tau}}
     e^{-\frac{(y-x-\drift_Q\tau)^2}{4D\tau}-\alpha \tau}
     \left( 1 +\mathcal{O}(\tau^2)\right)  \ .
\end{equation}
Much of the calculation of the entropy production 
 follows the procedure in 
 Secs.~\ref{sec:DDparticle} and \ref{sec:Bpartring} to be detailed further below. To this end, we also need 
 \begin{align}
     \lim_{\tau\to 0}\frac{\plaind}{\plaind\tau}W(x\to y, 1\to 2;\tau)&=\dot W(x\to y, 1\to 2)=\alpha\delta(x-y)\nonumber\\
     &=\dot W(x\to y,2\to 1)\,.
     \end{align}
     As far as processes are concerned that involve a change of particle mode, therefore only the transition rates enter, not diffusion or drift. Given a uniform stationary spatial distribution of particles of any mode, mode changes between two modes cannot result in a sustained probability current, even when the switching rates differ, 
     \begin{equation}
         \left(P_1 \dot{W}(1\to 2) - P_2 \dot{W}(2\to 1)\right)
         \ln\left(
         \frac{P_1 \dot{W}(1\to 2)}{P_2 \dot{W}(2\to 1)}
         \right)
         =0
     \end{equation}
     for $P_1 \dot{W}(1\to 2)=P_2 \dot{W}(2\to 1)$ at stationarity as in the process discussed in 
     \Sref{sec2state}.
     A probability current and thus entropy production can occur 
     when different particle modes result in  a different distribution, \Sref{OU},
     or when mode switching between more than two modes results in a current in its own rights, Secs.~\ref{sec:sec3state} and \ref{sec:switchdiff}.
     
     Since the full time-dependent density is beyond the scope of the present work, we calculate entropy flow and production at stationary on the basis of a natural extension of 
     \Erefs{external_en_prod_and_internal_en_prod},
     \eref{continuum_en_flow}
     and \eref{Se_cont_easy} to a mixture of discrete and continuous states
\begin{align}
\phantom{=}
&-\lim_{t\to\infty}\entropyFlow(t)=
\lim_{t\to\infty}\entropyProduction(t)\\
&=\sum_{Q,R\in\{1,2\}}\int^L_0 \plaind x \plaind y P_Q(x,t)\dot W(x\to y,Q\to R)\lim_{\tau\to 0}\ln\left(\frac{W(x\to y,Q\to R;\tau)}{W(y\to x, R\to Q;\tau)} \right)\nonumber\\
&=\frac{\drift_1^2+\drift_2^2}{2D} \label{eq:rnt_steady_si}
\end{align}
which immediately follows from Secs.~\ref{sec:DDparticle}
and \ref{sec:Bpartring}, as the stationary density is constant, $P_Q=P_R=1/(2L)$, and only $Q=R$ contribute, with
\begin{equation}
  \lim_{\tau\to0}\ln\left(\frac{W(x\to y,1\to 2;\tau)}{W(y\to x,2\to 1;\tau)}\right) =0 \ .
\end{equation}

If the drifts are equal in absolute value 
$|\drift_1|=|\drift_2|=\drift$, then we recover the entropy production of a simple 
drift-diffusive particle,
$\entropyProduction={\drift^2}/{D}$. This is because we can think of run-and-tumble as a drift-diffusion particle that changes direction instantly. Since changing the direction produces no entropy, the total entropy production rate should be the same as a drift-diffusion particle. 
The entropy production  can alternatively be derived via (\ref{eq:tot_en_prod}) by computing $\entropyProduction = \int \plaind x \left( j_1^2/(DP_1) + j_2^2/(DP_2)\right)$ with the steady-state currents stated above.

\subsection{Switching diffusion process on a ring}
\label{sec:switchdiff}
\begin{figure}[h]
\centering
\includegraphics[width=0.4\textwidth]{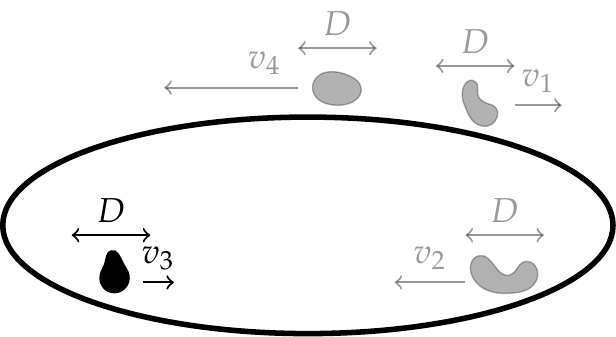}
\caption{\flabel{F13} Switching diffusion process on a ring $x \in [0,L]$ in continuous time. A switching diffusion process involves a stochastic switching between $M$ modes characterised by an identical diffusion constant $D$ but distinct drifts $\drift_i$ ($i=1,2,\dots,M$). The marginal switching dynamics are characterised as an $M$-state Markov process with transition rates $\alpha_{ij}$ from mode $i$ to mode $j$.}
\end{figure} 

The dynamics of a one-dimensional run-and-tumble particle discussed above can be readily generalised to the so called switching diffusion process \cite{Yang2018Jul} by allowing for an extended set $\{\drift_i\}$ of drift modes $i=1,\dots,M$, \fref{F13}. The corresponding Langevin equation for the particle position on a ring $x \in [0,L]$ is almost identical to that of run-and-tumble, namely $\dot{x} = \drift(t) + \sqrt{2D}\xi(t)$, with the exception that the process $\drift(t)$ is now an $M$-state Markov process. In the general case, a single switching rate $\alpha$ is thus not sufficient and the full transition rate matrix $\alpha_{ij}$ needs to be provided. In this formulation, the run-and-tumble dynamics Sec.~\ref{sec:RnTring} correspond to the choice $M=2$ with symmetric rates $\alpha_{12}=\alpha_{21}=\alpha$. Defining $P_i(x,t)$ as the joint probability that at time $t$ the particle is at position $x$ and in mode $i$, thereby moving with velocity $\drift_i$, the system (\ref{eq:system_FP_rnt}) of Fokker-Planck equations generalises to 
\begin{equation}
    \partial_t P_i(x,t)=-\partial_x[(\drift_i-D\partial_x)P_i(x,t)] + \sum_j P_j(x,t) \alpha_{ji}  
\end{equation}
where the transmutation rates $\alpha_{ij}$ from mode $i$ to mode $j$ are assumed to be independent of position. To ease  notation we use the convention $\alpha_{jj} = - \sum_{i\neq j}\alpha_{ji}$. For non-vanishing diffusion constant, the stationary solution is uniform for all modes and given by $\lim_{t \to \infty} P_i(x,t) = \weight_i/ L$, where $\weight_i$ is the $i$th element of the eigenvector $\zvec$ satisfying $\sum_j \weight_j=1$ and the eigenvalue relation $\sum_j \weight_j \alpha_{ji} = 0$, which we assume to be unique for simplicity.

The calculation of the steady-state entropy production follows very closely that of run-and-tumble presented above. The conditional transition probabilities including up to one transmutation event read to leading order
\begin{equation}
\label{eq:propagators_swithdiff}
W(x\to y, i \to j;\tau)=
    \begin{cases}
     \frac{e^{\alpha_{ii}\tau}}{\sqrt{4\pi D \tau}} \exp{-\frac{\left(y-x-\drift_i \tau \right)^2}{4D \tau }}
     \left(1+\OC(\tau^2)\right)
     &\mbox{for } i = j \\
     \frac{\alpha_{ij}}{2(\drift_i-\drift_j)}\left[\text{erf}\left(\frac{x-y+ \drift_i\tau}{\sqrt{4D\tau}}\right)-\text{erf}\left(\frac{x-y+ \drift_j \tau}{\sqrt{4D\tau }}\right)\right] 
     \left(1+\OC(\tau^2)\right)
     &\mbox{for } i \neq j ~,
    \end{cases}
\end{equation}
so that 
\begin{equation}
\label{eq:tder_propag_switchdiff}
    \lim_{\tau \to 0} \frac{\plaind}{\plaind\tau} W(x \to y, i \to j;\tau) = 
    \begin{cases}
        D \partial_y^2 \delta(y-x) - \drift_i \partial_y \delta(y-x) + \alpha_{ii} \delta(y-x) &\mbox{for } i = j \\
        \alpha_{ij} \delta(y-x) &\mbox{for } i \neq j ~.
    \end{cases}
\end{equation}
We could perform the calculation of the entropy production using the procedure of  \sref{DDparticle} rather than drawing on the operator for $i = j$, which, however, is used in the following for convenience, see  \sref{Bpartring}.
Substituting (\ref{eq:propagators_swithdiff}) and (\ref{eq:tder_propag_switchdiff}) into (\ref{eq:Si_cont_easy}) and assuming steady-state densities, we arrive at
\begin{align}
    \lim_{t \to \infty} \entropyProduction(t) =& - \lim_{t \to \infty} \entropyFlow(t) \nonumber \\
    =& \int_0^L \plaind x \plaind y \ \sum_i \frac{\weight_i}{L} \left( D \partial_y^2 \delta(y-x) - \drift_i \partial_y \delta(y-x) + \alpha_{ii} \delta(y-x) \right) (y-x) \frac{\drift_i}{D} \nonumber \\
    &+ \int_0^L \plaind x  \plaind y \ \sum_{i, j \neq i} \frac{\weight_i}{L} \alpha_{ij} \delta(y-x) \ln \left( \frac{\alpha_{ij}}{\alpha_{ji}} \right) ~,
\end{align}
where we have used \Eref{tder_propag_switchdiff} in the operators containing the $\delta$-functions and \Eref{propagators_swithdiff} in the logarithms. The term $\ln(\alpha_{ij}/\alpha_{ji})$ is obtained by the same expansion as used in \Eref{RnT_trans_expanded},  \sref{RnTring}.
Both terms contributing to the entropy production above are familiar from previous sections: the first is a sum over the entropy production of $M$ drift-diffusion processes with characteristic drift $\drift_i$, Sec.~\ref{sec:Bpartring} without potential, weighted by the steady-state marginal probability $\weight_i$ for the particle to be in state $i$; the second is the steady-state entropy production of an M-state Markov process with transition rate matrix $\alpha_{ij}$, which reduces to \Eref{external_en_prod_and_internal_en_prod} after integration. Carrying out all integrals, we finally have
\begin{equation}
\label{eq:en_prod_switch_diff}
    \lim_{t \to \infty} \entropyProduction(t) = \lim_{t \to \infty} -\entropyFlow(t) = \sum_i \weight_i \frac{\drift_i^2}{D} + \frac{1}{2} \sum_{i,j} (\weight_i \alpha_{ij} - \weight_j \alpha_{ji}) \ln \left( \frac{\alpha_{ij}}{\alpha_{ji}} \right) ~.
\end{equation}
Unlike run-and-tumble, \sref{RnTring}, the transmutation process in switching diffusion does in general contribute to the entropy production for $M > 2$, since the stationary state generally does not satisfy detailed balance.
However, contributions to the total entropy production originating from the switching and those from the diffusion parts of the process are effectively independent at steady state, as only the stationary marginal probabilities $\weight_i$ of the switching process feature as weights in the entropy production of the drift-diffusion. Otherwise the parameters characterising the two processes stay separate in \Eref{en_prod_switch_diff}. Further, the drift-diffusion contributions of the form $\drift_i^2/D$ are invariant under the time-rescaling $\alpha_{ij} \to T \alpha_{ij}$. This property originates from the steady-state distributions $P_i(x)$ being uniform and would generally disappear in a potential, \sref{OU}. 

\section{Discussion and concluding remarks}
In this work we calculate  the rate of entropy production within Gaspard's framework \cite{gaspard_time-reversed_2004} from first principles in a collection of paradigmatic processes, encompassing both discrete and continuous degrees of freedom.  
Based on the Markovian dynamics of each system, where we can, we derive the probability
distribution of the particle (or particles) as a function of time $P(x,t)$ from Dirac or Kronecker-$\delta$ initial conditions $P(x,0)=\delta(x-x_0)$, 
from which  the transition probability $W(x\to y;\tau)$ follows straightforwardly. In some cases, we determine only the stationary density and the (short-time) propagator $W(x\to y;\tau)$ to leading order in $\tau$. We then use \Eref{external_en_prod_and_internal_en_prod} for discrete systems
or Eqs.~(\ref{eq:Si_cont_all}) and (\ref{eq:Se_cont_all}) for continuous systems to calculate the time-dependent entropy production.
We set out to give concrete, exact results in closed form, rather than general expressions that are difficult to evaluate, even when we allowed for general potentials in Sec.~\ref{sec:Bpartring}.
In summary, the ingredients that are needed to calculate the entropy production 
in closed form in the present framework are: a) the probability (density) $P(x,t)$ to find the system in state $x$ ideally as a function of time $t$ and b) the propagator $W(x \to y;\tau)$, the probability (density) that the system is found at a certain state $y$ after some short time $\tau$ given an  initial state $x$. If the propagator is known for any time $\tau$, it can be used to calculate the probability $P(x,t;x_0) = W(x_0 \to x;t)$ for some initial state $x_0$. However, this full time dependence is often difficult to obtain. The propagator is further needed in two forms, firstly $\lim_{\tau \to 0} {\partial_\tau} W(x\to y;\tau)$ when it is most elegantly written as an operator in continuous space, and secondly $\lim_{\tau \to 0} \ln (W(x\to y;\tau)/W(y\to x;\tau))$. 

For completeness, where feasible, we have calculated 
the probability current
$j(x,t)$
in continuous systems at position $x$. The mere presence of such a flow indicates broken time-reversal symmetry and thus non-equilibrium.
Our results on the 
discrete systems (\sref{sec2state} to \ref{sec:RWring}) 
illustrate two important aspects of entropy production. First, the need of a 
probability flow $P_A \dot{W}(A\to B) - P_B \dot{W}(B \to A)$ between states: in the two-state system \sref{sec2state}
there are no transition
rates $\alpha$ and $\beta$ such that there is a sustained probability flow 
and therefore, the system inevitably relaxes to equilibrium.
However, in the three-state system \sref{sec3state} the transition rates can
be chosen so that there is a perpetual flow $(\alpha - \beta)/3$ between any two states 
and therefore there is entropy
production not only during relaxation but also at stationarity. Hence, we can ascertain these as non-equilibrium steady states in the long time limit due to the non-vanishing rate of internal entropy production. 
Uniformly distributed steady states can be far from equilibrium as a rigorous analysis on the basis of the microscopic dynamics reveals, although an effective dynamics may suggest otherwise.

Second, we see how the extensivity of entropy production arises in the 
$N$-particle  systems (Secs.~\ref{sec:Ndist}, \ref{sec:N2p} and \ref{sec:Nindepd}), independently
of whether the particles are distinguishable or not. We therefore conclude that
the number of particles in the system must be accounted for when calculating
the entropy production, and doing otherwise will not lead to a correct
result. This is sometimes overlooked, especially when using effective theories.
In the continuous systems (\sref{DDparticle} to \ref{sec:Bpartring}), 
which involve a drift $\drift$ and a diffusion constant $D$,
we always find the contribution $\drift^2/D$ to the entropy production emerging
one way or another. Moreover, in the case of drift-diffusion on the real line 
(Sec.~\ref{sec:DDparticle}) we find that the contribution due to the relaxation of the system
$1/(2t)$ is independent of any of the system parameters.

Finally, we have studied two systems 
(\sref{RnTring} and \ref{sec:switchdiff}) where the state space has a discrete 
and a continuous component. The discrete component corresponds to the 
transmutation between particle species, \ie their mode of drifting, whereas the continuous component corresponds to the particle motion.
We find that both processes, motion and transmutation, contribute to the entropy
production rate essentially independently since any term that combines both processes is a 
higher-order term contribution in $\tau$, and therefore vanishes in the limit
$\tau\to0$.

This work has applications to the field of active particle systems, where
particles are subject to local non-thermal forces. In fact, the systems studied in sections
\ref{sec:sec3state} and
\ref{sec:RWring} --
\ref{sec:switchdiff}        
are prominent examples of active systems. 
We have shown that their entropy production crucially relies on the microscopic
dynamics of the system, which are captured by the Fokker-Planck equation 
(or the master equation for discrete systems)
and its solution.
However, in interacting many-particle systems, such a description is not available in general.
Instead, we may choose to use the Doi-Peliti formalism \cite{Doi:1976,Peliti:1985,TaeuberHowardVollmayr-Lee:2005,SmithKrishnamurthy:2018,BordeuETAL:2019,LazarescuETAL:2019,PauschPuressner:2019,Garcia-Millan:2020,Garcia-MillanPruessner:2020} to describe the system, since it provides a 
systematic approach based on the microscopic dynamics and which retains the particle entity.


\acknowledgments{ The authors would like to thank Letian Chen, Greg Pavliotis and Ziluo Zhang for discussions and kind advice. The authors gratefully acknowledge Kin Tat (Kenneth) Yiu's much earlier, related work \cite{KTYiu:2017}.}

\conflictsofinterest{None.} 
\bibliography{bib}{}
\authorcontributions{Formal analysis, Luca Cocconi, Rosalba Garcia-Millan, Zigan Zhen and Bianca Buturca; Supervision, Gunnar Pruessner; Writing – original draft, Luca Cocconi, Rosalba Garcia-Millan, Zigan Zhen and Bianca Buturca; Writing – review \& editing, Gunnar Pruessner.}
\end{document}